\documentclass[preprint,12pt]{elsarticle}

\usepackage{amsmath}
\usepackage{amsfonts}
\usepackage{amssymb}
\usepackage{graphicx}
\usepackage{latexsym}
\usepackage{mathtools}

\usepackage{floatrow}
\newfloatcommand{capbtabbox}{table}[][\FBwidth]

\def \semb#1{{\ldbrack #1 \rdbrack}}

\newcommand{\ar}[1]{\xrightarrow{#1}}

\newcommand{\dar}[1]{\xRightarrow{#1}}

\renewcommand{\P}{\mathcal{P}}
\newcommand{\Q}{\mathcal{Q}}
\newcommand{\G}{\mathcal{G}}
\newcommand{\R}{\mathcal{L}}

\newcommand{\N}{\mathcal{R}}

\newcommand{\TT}[1]{{\mathcal{TT}(#1)}}

\newcommand{\BT}[1]{{\mathcal{CT}(#1)}}

\usepackage[zed]{csp}

\newtheorem{definition}{Definition}
\newtheorem{lemma}{Lemma}
\newtheorem{theorem}{Theorem}

\newtheorem{proposition}{Proposition}

\begin{document}

\begin{frontmatter}

\title{Revisiting Timed Specification Theory II : Realisability}

\author{Chris Chilton}
\author{Marta Kwiatkowska}
\author{Xu Wang}

\address{Department of Computer Science, University of Oxford, UK}

\begin{abstract}
In this paper we present an assume-guarantee specification theory (aka interface theory from~\cite{henzinger-timedia}) for modular synthesis and verification of real-time systems with critical timing constraints. It is a further step of our earlier work~\cite{CKW12} which achieved an elegant algebraic specification theory for real-time systems endowed with the capability to freeze time. In this paper we relinquish such (unrealisable) capability and target more realistic systems without the ability to stop time. 

In comparison with related works~\cite{henzinger-timedia,larsen-timedio}, we build our theory on a surprisingly simple framework of timed I/O automata enhanced with \emph{invariant/co-invariant distinction}, which, nevertheless, suffices to specify the timed assumption and guarantee of a component w.r.t. both \emph{safety} and \emph{bounded-liveness} requirements. When two specifications are parallel composed, the guarantee in one specification will be matched against the assumption in the other. Any mismatch gives rise to an occurrence of \emph{incompatibility error}. 


Our theory, in a combined \emph{process-algebraic} and \emph{reactive-synthesis} style,  provides the operations of parallel composition for system integration, logical conjunction/disjunction for viewpoint fusion and independent development, and quotient for incremental synthesis. 

We show that a substitutive refinement preorder, which is a coarsening of the pre-congruence in~\cite{CKW12}, constitutes the weakest pre-congruence preserving freedom of incompatibility errors. The coarsening requires a shift in the focus of our theory to a more game-theoretical treatment, where the coarsening constitutes a reactive synthesis game named \emph{normalisation} and is efficiently implementable by a novel local \emph{$\bot$-backpropagation} algorithm. 

Previously, timed concurrent games have been studied in~\cite{Asarin98,henzinger-timedia,AFHMS03}, where one of the key concern is the removal of time-blocking strategies by applying blame assignment~\cite{AFHMS03}. Our timed games also have the issue of time-blocking strategies, which may arise through the composition of specifications. However, due to our distinctively different formulation of timed games, we have found another elegant solution to the problem without blame assignment. Our solution utilises a second reactive synthesis game called \emph{realisation}, which is dual to normalisation and implementable by the dual local \emph{$\top$-backpropagation} algorithm. 

Based on the timed game formulation and as a further step to previous works, we also study the \emph{composition of synthesis games} under different operators, e.g. the distributivity of realisation over conjunction, which arises through the composition of specifications, and which can also be usefully exploited as a theoretical foundation for the compositional synthesis~\cite{Filiot10} of timed processes.

Utilising such knowledge, we achieve the complete operational definition to all the composition operators (on specifications) and prove the weakest congruence result by applying the timed strategies semantics on the set of operators.

\end{abstract}

\begin{keyword}
timed automata \sep timed interfaces \sep specification theory \sep assume/guarantee verification\sep reactive/controller synthesis \sep weakest congruence \sep substitutive refinement  \sep conjunction \sep quotient \vspace*{-4pt}

\end{keyword}

\end{frontmatter}

\section{Introduction}
\label{sec:intro}

Modular synthesis and verification of quantitative aspects (e.g. real-time, probability, reward, etc.) of computational and physical processes (e.g. cyber-physical systems) is an important research topic. For instance, \cite{Benveniste} gives a general discussion and motivation of the modular approach to quantitative system design. In this programme of quantitative study, a specification of components consists of a combination of quantitative assumption and quantitative guarantee. One of the crucial criteria for the success of such a programme lies in a \emph{unified core theory}, to which only \emph{minimal and additive extensions} are required for addressing the different aspects, so that the amalgamation of the extensions does not entail overwhelming technical complications. 

As one step of the programme, this paper targets component-based development for real-time systems with critical timing constraints, such as embedded system components, the middleware layer and asynchronous hardware. We propose a complete timed specification theory using a framework of  \emph{minimal extension of timed automata}. 

The framework provides the operations of parallel composition for examining the structural behaviour of systems, logical conjunction/disjunction for viewpoint fusion and independent development, as well as quotient for incremental synthesis. 

The refinement relation is defined relative to the notion of \emph{incompatibility error}. That is, parallel composition incurs the matching up of the assumption and guarantee from different components. Any AG mismatch generates an incompatibility error (denoted by $\bot$) in the composed system. Refinement thus means error-free substitutivity: there is no context in which replacing a component by a refinement will introduce further incompatibility error.\footnote{Note that the existence of incompatibility errors does not mean that the composed system is un-usable; an environment can still usefully exploit the system by only exercising the parts of its behaviours insulated from the incompatibility errors, as has been well explained in~\cite{henzinger-timedia}.}


Previously, based on the framework, \cite{CKW12} introduced a compositional linear-time specification theory for real-time systems, where the substitutive refinement is the weakest pre-congruence preserving incompatibility errors (for the four operations), and characterisable by a finite trace semantics. A key novelty of~\cite{CKW12} lies in the introduction of an explicit \emph{timestop} operation  (denoted by $\top$) that halts the progress of the system clock.  

Equipped with timestop, an environment of~\cite{CKW12} 1) can tell two components apart by observing not only the occurrence of incompatibility errors but also the timing difference in such occurrences, and 2) can steer any component away from incompatibility errors no matter how error-prone it is. Thus, it gives rise to a finest congruence over a set of fully defined operators (esp. conjunction and quotient) as well as a greatly simplified theory.


While timestop is appropriate for a restricted class of applications, such as embedded systems and circuit design~\cite{MTCMR00}, there are cases where the operation of stopping the system clock is neither meaningful nor implementable. Similar observations have also been made in the works on concurrent timed games~\cite{Asarin98,henzinger-timedia,AFHMS03}, where there is no explicit timestop operation but the use of implicit timestop by time-blocking strategies is considered unrealistic for winning games. Thus, it is desirable to consider systems without explicit or implicit timestops, which we call \emph{realisable systems}.

For realisable systems, components, not substitutively-equivalent according to~\cite{CKW12}, can become equivalent under realisability. This is a consequence of the environment losing the power to observe the timing difference in error occurrences (see the example in Figure~\ref{fig:distinct1}). Thus, we need a new substitutive refinement preorder, which is a coarsening of the pre-congruence in~\cite{CKW12}. 

To best characterise the coarsening, our theory needs a shift in focus to a more game-theoretical treatment\footnote{In contrast, our early work~\cite{CKW12} is based predominantly on a process-algebraic and trace-theoretical framework, where the timed game part plays only the supportive role for providing a general setting to timed strategies semantics.}, where the coarsening constitutes a reactive synthesis game called \emph{normalisation}, and is efficiently implementable by a novel local \emph{$\bot$-backpropagation} algorithm which repeatedly removes incompatibility errors from a system. The \emph{$\bot$-backpropagation} algorithm is strictly more aggressive (i.e. classifying more states as winning states) than the classical timed reactive synthesis algorithms~\cite{Asarin98,Cassez05} and is crucial for our weakest congruence results. 

Furthermore, similar to timed concurrent games~\cite{henzinger-timedia,AFHMS03}, where one of the key concern is the removal of time-blocking strategies by applying blame assignment, it is also crucial in our framework to remove timestopping behaviours since specification composition (e.g. conjunction and quotient) may generate new unrealisable behaviours. However, unlike~\cite{henzinger-timedia,AFHMS03}, our framework does not use blame assignment to remove unrealisable behaviours. Rather, we have found a different elegant solution based on a dual reactive synthesis game to normalisation called \emph{realisation}, largely thanks to our different formulation of timed games. Realisation can be efficiently implemented by the dual \emph{$\top$-backpropagation} algorithm.



Furthermore, unlike previous works on timed concurrent games~\cite{Asarin98,henzinger-timedia,AFHMS03,Cassez05,larsen-timedio}, which mostly concentrating on studying a single game, our work also studies the composition of games under different operators. That is, each specification is embedded with a pair of synthesis games. When specifications are composed, we need to understand how the synthesis games interact or interfere with one another across specification boundary and how should we define the composition of such games correctly. This will form a basis for both the compositional synthesis of timed processes and the full operational definition of specification composition operators.

Finally, some further contributions of our theory lie in 1) the process-algebraic techniques of deriving process composition operation from state composition operation via \emph{state-to-process lifting}, enabling the transfer of algebraic properties from the state composition level to the process composition level, 2) the robust and intuitive \emph{timed-strategies} characterisation of the refinement and operators, which serves as a simple correctness proof to the operator definitions, 3) the linear-time (i.e. double trace sets) characterisation of the refinement and operators, which supports the explicit separation of assumption and guarantee and interfaces well with automata and learning techniques, and 4) the elegant minimal extension of timed automata that can distinguish, for the first time, the roles of I/O transition guards and invariant/co-invariant as specifying resp. timed safety/liveness assumptions/guarantees, thus making our TIOAs an appealing model for practical application of timed AG reasoning. 



\paragraph{Outline} Section~\ref{frame} introduces a minimal extension of timed automata as our formal framework, i.e. timed I/O automata (TIOA) and timed I/O transition systems (TIOTS). Based on TIOTSs, we introduce 1) the $\bot$ state and the auto-$\bot$/semi-$\bot$ states as incompatibility errors in closed systems and open systems resp., and 2) the auto-$\top$ and semi-$\top$ states as explicit and implicit timestop. Based on $\top$- and $\bot$- completed TIOTSs, we define the parallel composition operator using the state-to-process lifting technique. 

Section~\ref{sec:game} introduces our formulation of timed I/O games, consisting of three players, system, environment and coin. Then we define game rules and strategies and show that the parallel composition of specifications can be reduced to strategy composition. Finally we define refinement as error-free substitutivity and give the corresponding strategy characterisation via a so-called determinisation procedure that converts imperfect-information games into perfect information games.

Section~\ref{sec:realise} introduces the concept of realisable specifications as well as the coarsened refinement. Then, we introduce the timed synthesis game called \emph{normalisation} and shows that auto-$\bot$/semi-$\bot$ states are localised version of $\bot$-winning states in such games. Finally, using the normalised strategies, we illustrate what the expected semantics is for the operators like conjunction, disjunction and quotient.


Section~\ref{sec:opsem} gives the operational definition of the operators using a combined process-algebraic and reactive-synthesis style. We first give the process-algebraic definitions (i.e. state-to-process lifting) for the restricted cases when operands are all normalised, and show 1) that the composition under conjunction and quotient may generate new unrealisable (i.e. time-blocking) strategies that is removable by another reactive-synthesis game called \emph{realisation}, and 2) that semi-$\top$/auto-$\top$ states are localised version of the \emph{$\top$-winning states} for the realisation game.

Then we give the reactive-synthesis operational definitions for the general cases when specifications are not normalised. We study how the synthesis games interfere with each other across the specification boundary under different operators. We prove results like the distributivity of normalisation/realisation over operations like conjunction, quotient, and determinisation. 

Finally, Section~\ref{sec:exam} uses a case study to illustrate how we can use our novel backpropagation to synthesise controllers that can steer a component away from undesirable behaviours. Related work is considered in Section~\ref{sec:comp}, while we conclude and suggest future work in Section~\ref{sec:concl}.

\section{Minimal TA Extension for Timed Specifications}
\label{frame}

In this section we introduce our timed framework, i.e. \emph{timed I/O automata} (TIOA) and \emph{timed I/O transition systems} (TIOTS). Our framework has significant differences from the timed models defined by previous works~\cite{Kaynar,henzinger-timedia,larsen-timedio}. The distinction mostly lies in that our models are specially designed to support the \emph{mixed assume/guarantee specifications of components}. That is, given a component, we specify both its system guarantee and environmental assumption, which are combined and mixed to be represented by a single automata. In this respect our specifications are similar to timed interfaces proposed by~\cite{henzinger-timedia}.





The origin of our framework appeared earlier in our work~\cite{CKW12}.  However, the version presented in this section contains important technical extension as well as presentation improvements.





\subsection{Timed I/O Automata}\label{sec:time}

Specifications in our theory are modelled by timed I/O transition systems, which can be compactly represented as timed I/O automata under certain restrictions. 


\paragraph{Clock constraints} Given a set $X$ of real-valued clock variables, a \emph{clock constraint} over $X$, $cc: CC(X)$, is a boolean combination of atomic constraints of the form $x \bowtie d$ and $x - y \bowtie d$, where $x,y \in X$, $\bowtie \in \{\leq,<, =, >,\geq\}$, and $d \in \mathbb{N}$.

\begin{definition}
A \emph{timed I/O automaton (TIOA)} is a tuple $(C, I, O, L, l^0, AT,$ $ Inv, coInv)$, where:

\begin{itemize}
\item $C \subseteq X$ is a finite set of clock variables (ranged over by $x,y$, etc.)

\item $A = I \uplus O$ is a finite alphabet (ranged over by $a,b$, etc.) consisting of the inputs $I$ and outputs $O$

\item $L$ is a finite set of \emph{locations} (ranged over by $l,l'$, etc.)

\item $l^0 \in L$ is the \emph{initial location}

\item $AT \subseteq L \cross CC(C) \cross A \cross 2^{C} \cross L$ is a set of \emph{action transitions}

\item $Inv : L \fun CC(C)$ and $coInv : L \fun CC(C)$ assign \emph{invariants} and \emph{co-invariants} to states, each of which is a downward-closed clock constraint.
\end{itemize} 
\end{definition}

In the rest of the paper we use $l \ar{g,a,rs} l'$ as a shorthand for $(l, g, a, rs, l') \in AT$. $g: CC(C)$ is the enabling guard of the transition, $a \in A$ the action, and $rs$ the subset of clock variables to be reset. 

Our TIOAs are an extension of timed automata that distinguish \emph{input from output} and \emph{invariant from co-invariant}. They are designed for the assume/guarantee specification of timed components, and can be regarded as a simplification of the timed interface automata of~\cite{henzinger-timedia}. 
In our framework, a specification is a combination of 
the timing assumptions made by the component on the inputs issued by the environment \emph{along with} the timing guarantees provided by the component on its outputs. Specifically:

\begin{itemize}
\item Guards on output transitions express \emph{safety timing guarantees}. The component guarantees that an output will only be fired at a point in time when it is allowed by a guard. 

\item Guards on input transitions express \emph{safety timing assumptions}. The component assumes that the environment will only issue an input at a time when it is allowed by a guard. 

\item An invariant (at a location) expresses \emph{liveness timing guarantees}. The system guarantees that some output will be fired before the time bound specified by the invariant has been exceeded. 

\item A co-invariant expresses \emph{liveness timing assumptions}. The component assumes that the environment will issue some input before the time bound specified by the co-invariant has been exceeded.
\end{itemize}

\begin{figure}[t]
\begin{center}
\includegraphics[width=\textwidth]{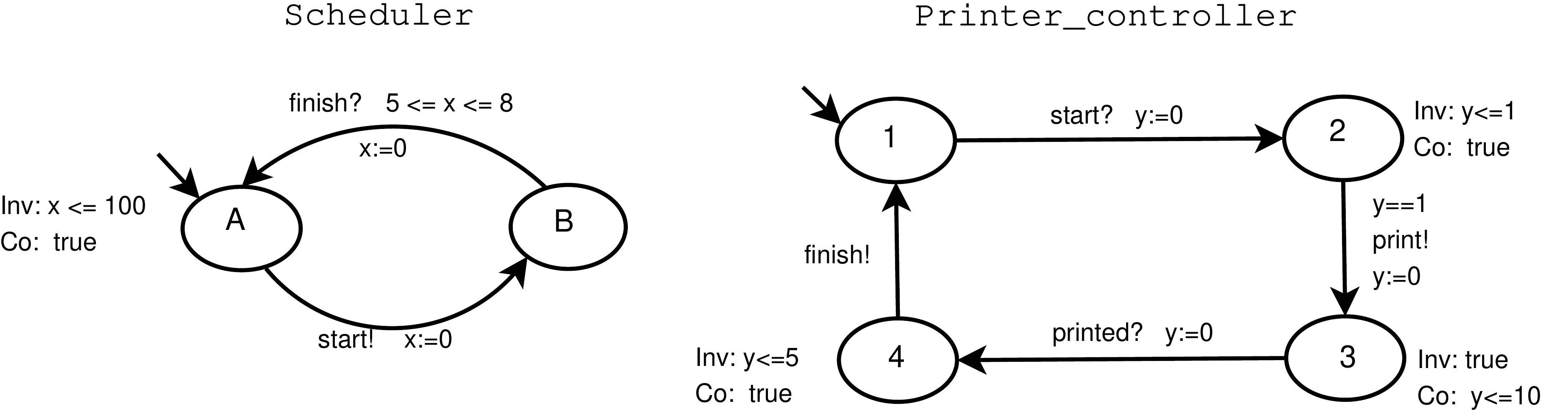}
\end{center}
\caption{Job scheduler and printer controller.}
\label{fig:automata}
\end{figure}

\paragraph{Example} Figure~\ref{fig:automata} depicts TIOAs representing a job scheduler together with a printer controller. The invariant at location $A$ of the scheduler forces a \emph{bounded-liveness guarantee} on outputs in that location: as time must be allowed to progress beyond $x=100$, the $start$ action must be fired before $x$ exceeds $100$. After $start$ has been fired, the clock $x$ is reset to $0$ and the scheduler waits (possibly indefinitely) for the job to $finish$. In the case that the job does finish, the scheduler expects this to take place only at a time point satisfying $5\leq x \leq 8$ (i.e. \emph{safety assumption}).

The controller waits for the job to $start$, after which it will wait exactly $1$ time unit before issuing $print$ (forced by the invariant $y\leq 1$ on state $2$ and the guard $y=1$ on the $print!$ transition, acting together as a combined liveness and \emph{safety guarantee}). Then, the controller requires the printer to acknowledge the job as having been $printed$ within $10$ time units (i.e. co-invariant $y\leq 10$ in state $3$ acting as \emph{bounded-liveness assumption}). After receiving the acknowledgement, the controller must indicate to the scheduler, within $5$ time units, that the job has $finish$ed.

\subsection{Timed I/O Transition Systems}\label{sec:timetrans}

Formally, the semantics of TIOAs are given by a minimal extension of timed transition systems, which are a special class of infinite labelled transition systems enhanced with two distinguished states $\top$ and $\bot$.

\begin{definition}
A \emph{timed I/O transition system (TIOTS)} is a tuple $\P= \langle I, O, S,$ $ s^0, \rightarrow \rangle$, where $I$ and $O$ are the input and output actions respectively,  $S = (L \times \mathbb{R}^C) \uplus \{\bot, \top\}$ is a set of states, $s^0 \in S$ is the designated initial state, and $\rightarrow \subseteq S \cross (I \uplus O \uplus \mathbb{R}^{>0}) \cross S$ is the action and time-labelled transition relation.
\end{definition}

\paragraph{Plain states} A \emph{clock valuation} over $C$ is a map $t$ that assigns to each clock variable $x$ in $C$ a real value from $\mathbb{R}^{\geq 0}$. A state of the TIOTS is a pair drawn from $L\times \mathbb{R}^{C}$ (i.e. the location and clock valuation pair), which we refer to as the set of \emph{plain states}. 

In addition, we introduce two special states $\bot$ and $\top$. These can be explained from a game-theoretic perspective. $\bot$ represents the violations of the assumptions on the environment, while $\top$ represents the violations of the guarantees by the system. Therefore, the system 
tries to avoid $\top$, while the environment tries to avoid $\bot$. The trivial TIOTS with $\top$ (resp. $\bot$) as the initial state is called the \emph{$\top$-TIOTS} (resp. \emph{$\bot$-TIOTS}). 

\paragraph{Notation} In the rest of the paper we use $p,p',p_i$ to range over plain states $P = L\times \mathbb{R}^{C}$ while $s,s',s_i$ range over $S$. Furthermore we define $tA= I \uplus O \uplus \mathbb{R}^{>0}$ to be the set of \emph{timed actions}, $tI= I  \uplus \mathbb{R}^{>0}$ to be the set of \emph{timed inputs}, and $tO= O \uplus \mathbb{R}^{>0}$ to be the set of \emph{timed outputs}. Symbols like $\alpha$, $\beta$, etc. are used to range over $tA$. 

A \emph{timed trace} (ranged over by $tt, tt',tt_i$ etc.) is a finite mixed sequence of positive real numbers ($\mathbb{R}^{> 0}$) and visible actions such that \emph{no two numbers are adjacent to one another}. For instance, $\langle 0.33, a, 1.41, b, c, 3.1415 \rangle$ is a timed trace denoting the observation that action $a$ occurs at $0.33$ time units, then another $1.41$ time units elapse before the simultaneous occurrence of $b$ and $c$, which is followed by $3.1415$ time units of no event occurrence. The empty trace is denoted by $\epsilon$. An infinite timed trace is an infinite such sequence.

We use $l(tt)$ to indicate the duration of $tt$, which is obtained as the sum of all the reals in $tt$, and use $c(tt)$ to count the number of action occurrences along $tt$. Concatenation of timed traces $tt$ and $tt'$, denoted $tt\cat tt'$, is obtained by appending $tt'$ onto $tt$ and coalescing adjacent reals (summing them). For instance, $\langle a, 1.41 \rangle$ $\cat \langle 0.33, b, 3.1415 \rangle$ = $\langle a, (1.41 + 0.33), b, 3.1415 \rangle$ = $\langle a, 1.74, b, 3.1415 \rangle$. 

Prefix/extension are defined as usual by concatenation. We write $tt \upharpoonright tA_0$ for the projection of $tt$ onto timed alphabet $tA_0$, which is defined by removing from $tt$ all actions not inside $tA_0$ and summing up adjacent reals.

\paragraph{Determinism and Non-zenoness}
We say a TIOTS is \emph{deterministic} iff there is no ambiguous transition, i.e. $s \ar{\alpha} s' \wedge s \ar{\alpha} s''$ implies $s'=s''$. It is \emph{time additive} providing $p \ar{d_1+d_2} s'$ iff $p \ar{d_1} s$ and $s \ar{d_2} s'$ for some $s$. 

For a TIOTS $\P$, we use $p \dar{tt} p'$ to denote a finite execution starting from $p$ that produces trace $tt$ and leads to $p'$. Similarly, we can define infinite executions which produce infinite traces on $\P$. An infinite execution is \emph{zeno} iff the action count is infinite but duration is finite. 

We say a TIOTS $\P$ is \emph{non-zeno} providing no plain execution is zeno. $\P$ is \emph{strongly non-zeno} iff there exists some $k \in \mathbb{N}$ s.t., for all plain executions $p \dar{tt}p'$, it holds that $l(tt)=1$ 
implies $c(tt) \leq k$. Here, we say a finite or infinite execution is a \emph{plain execution} iff the execution only visits plain states.

\paragraph{Assumption on TIOTSs} We only consider non-zeno time-additive TIOTSs in this paper. For technical convenience (e.g. ease of defining time additivity and trace semantics), the definition of TIOTSs requires that $\top$ and $\bot$ are \emph{chaotic states}, i.e. a state in which the set of outgoing transitions are all self-loops, one for each $\alpha \in tA$. 

The strong non-zenoness is not an assumption of our theory. But with this additional requirement we can show that the synthesis and verification theory in this paper is fully automatable.

\subsection{From TIOAs to TIOTSs}

In this section we show how to derive a TIOTS that represents the semantics of a TIOA. 

\paragraph{$\top/\bot$ completion} We first introduce two semantics-preserving transformations on TIOTSs, which give an explicit representation for assumption and guarantee violations. The \emph{$\bot$-completion} of a TIOTS $\P$, denoted $\P^{\bot}$, adds an $a$-labelled transition from $p$ to $\bot$ for every $p \in P$ ($=L\times \mathbb{R}^{C}$) and $a \in I$ s.t. $a$ is not enabled at $p$.\footnote{$\bot$-completion will make a TIOTS \emph{input-receptive}, i.e. input-enabled at all states.} The \emph{$\top$-completion}, denoted $\P^{\top}$, adds an $\alpha$-labelled transition from $p$ to $\top$ for every $p \in P$ and $\alpha \in tO$ s.t. $\alpha$ is not enabled at $p$. This coincides with our game-based interpretation of $\top$ and $\bot$, since:

\begin{enumerate}
\item a disabled input at a plain state is represented as an input transition to $\bot$ (assumption violation)
\item a disabled output at a plain state is represented by an output transition from that state to $\top$ (guarantee violation)
\item a disabled delay is represented by a delay transition to $\top$ (guarantee violation).  
\end{enumerate}




The mapping of disabled delays to $\top$ looks surprising, since time is neither controlled by the system or environment. Our bias towards $\top$ is due to a decision made relating to urgency semantics. 

In classical semantics without I/O distinction, if a state has no delay transition enabled, then some action becomes 
urgent for firing. For I/O systems, if a state has no enabled delay transition, we have to choose either the inputs or the outputs (enabled at that state) to become urgent.

The above mapping of disabled delays to $\top$ implies we choose to make outputs urgent, since the pending $\top$ (guarantee violation) implies the system cannot let time pass and so must fire with urgency other transitions under its control (i.e. an output transition).

\paragraph{$\top$/$\bot$ removal} The inverse operations of $\top/\bot$ completion, called \emph{$\top/\bot$ removal}, are also semantic-preserving transformations. For instance, $\top$-removal removes all output and delay transitions from plain states to $\top$ in the TIOTSs.

\

We can now give the execution semantics of TIOAs in term of $\top/\bot$-removed TIOTSs, since it will make the mapping simpler. 

\paragraph{Clock valuation}
We say a clock valuation $t$ satisfies a clock constraint $cc$, written $t \in cc$, if $cc$ evaluates to true under valuation $t$. $t + d$ denotes the valuation derived from $t$ by increasing the assigned value on each clock variable by $d \in \mathbb{R}^{\geq 0}$ time units. $t[rs \mapsto 0]$ denotes the valuation obtained from $t$ by resetting the clock variables in $rs$ to $0$. Sometimes we use $0$ for the clock valuation that maps all clock variables to $0$. 

\begin{definition}
The semantic mapping of a TIOA $\P$ is a TIOTS $\langle I, O, S,\allowbreak s^0, \allowbreak\rightarrow \rangle$ with:
\begin{itemize}
\item set of states $S = (L \times \mathbb{R}^C) \uplus \{\bot, \top\}$
\item initial state $s^0 = \top$ providing $0 \notin Inv(l^0)$, $s^0 = \bot$ providing $0 \in  Inv(l^0) \wedge \neg coInv(l^0)$ and $s^0 = (l^0, 0)$ providing $0 \in  Inv(l^0) \wedge coInv(l^0)$,
\item a transition relation $\rightarrow \subseteq S \cross (I \uplus O \uplus \mathbb{R}^{>0}) \cross S$ being the smallest (time-additive) relation such that:
\begin{enumerate}
\item $\top$ and $\bot$ are chaotic states,
\item If $l \ar{g,a,rs} l'$, $t'= t [rs \mapsto 0]$, $t \in Inv(l) \wedge coInv(l) \wedge g$, then:
\begin{enumerate}
\item \emph{plain action:} $(l, t) \ar{a} (l', t')$ providing $t' \in Inv(l') \wedge coInv(l')$
\item \emph{magic action:} $(l, t) \ar{a} \top$ providing $t' \in \neg  Inv(l')$ and $a \in I$
\item \emph{error action:} $(l, t) \ar{a} \bot$ providing $t' \in Inv(l') \wedge \neg coInv(l')$ and $a \in O$.
\end{enumerate}
\item \emph{plain delay:} $(l, t) \ar{d} (l, t+d)$ if $t,t+d \in Inv(l) \wedge coInv(l)$
\item \emph{time-out delay:} $(l, t) \ar{d} \bot$ if $t \in Inv(l) \wedge coInv(l)$ 
and $t+d \in Inv(l) \wedge \neg coInv(l)$.\footnote{Note that by time additivity and the chaotic nature of $\bot$: $p \ar{d} \bot$ implies $p \ar{d'} \bot$ for all $d' \geq d$.}

\end{enumerate}
\end{itemize}
\end{definition}

\noindent
In TIOAs we do not have explicit $\top$ and $\bot$. This is because we interpret a configuration $(l,t)$ as $\top$ if  $t$ violates the invariant in location $l$ and we interpret a configuration $(l,t)$ as $\bot$ if $t$ violates the co-invariant in location $l$ (while the invariant holds). The two types of configurations are collectively called \emph{illegal configurations}.  Sometimes we simply represent a location with true as the invariant and false as co-invariant by $\bot$. Dually, we have a $\top$ location. 

The TIOTS attempts to track the configuration of the TIOA, and directly maps the illegal configurations to $\top$ and $\bot$. Furthermore, our TIOTS does not contain transitions that are $\top$/$\bot$-removable. As a consequence, only output and delay transitions go to $\bot$ and only input transitions go to $\top$.

Note that our interpretation gives \emph{priority to the invariant} (cf the occurrences of the condition $Inv \wedge \neg coInv$ in the above definition). If a delay exceeds the invariant bound before exceeding the co-invariant bound, the delay transition goes to $\top$, which is modelled as a disabled transition; if a delay exceeds the co-invariant bound before exceeding the invariant bound, the delay transition goes to $\bot$ (i.e. \emph{time-out delay}). However, if a delay exceeds both bounds simultaneously, the delay transition goes to $\top$ (i.e. as a disabled transition). 

\subsection{Parallel composition}

In the rest of the paper, we will develop our theory on top of TIOTSs, which are endowed with a richer repertoire of semantic machinery.\footnote{Furthermore, we will not restrict ourselves to TIOTSs mapped from TIOAs.} In particular, we will use $\top/\bot$-completed TIOTSs extensively, since the nice duality possessed by $\top/\bot$-completed TIOTSs can simplify our presentation a lot. But, from time to time, we will also use $\top/\bot$-removed TIOTSs or even $\top/\bot$-free TIOTSs, because, without $\top$ and $\bot$, the TIOTSs are essentially classical I/O transition systems~\cite{Kaynar,WangKwiat07}, enabling us to tap into classical semantics.

Therefore, we will freely switch between the two levels of semantics in the sequel: $\top$/$\bot$-completed TIOTSs and $\top$/$\bot$-removed TIOTSs. Sometimes, when defining a new construct, the intuition is strong and clear on one level, but not on the other. So we will formulate the construct on the former and then extrapolate into the latter.



Let us start with the parallel composition operator, the most important operator in a specification theory. We will define the operator on top of $\top$/$\bot$-completed TIOTSs. But the intuition comes from the definitions with classical semantics.

\begin{figure}[t]
\begin{center}
\includegraphics[width=0.5\textwidth]{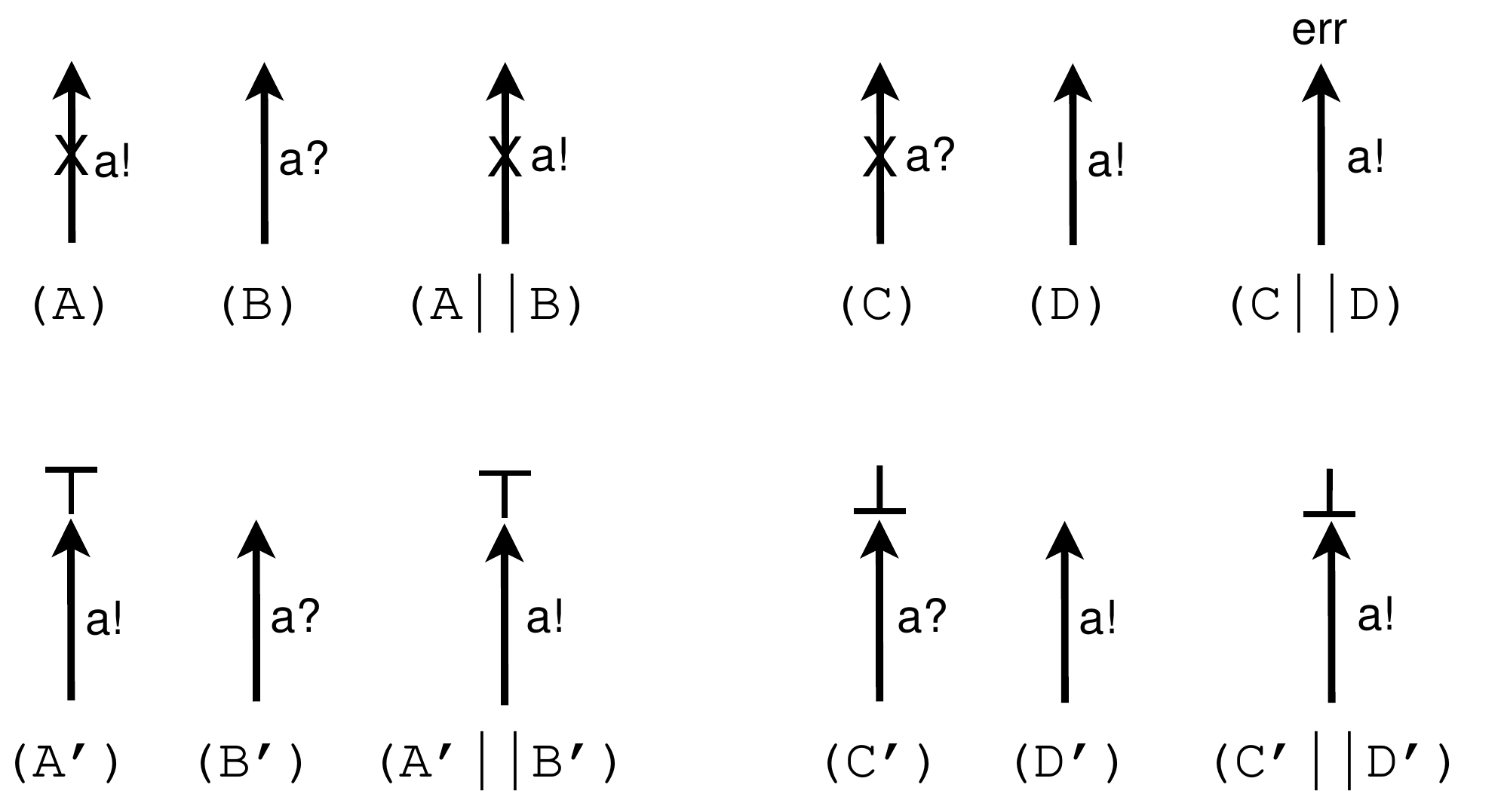}
\end{center}
\caption{Parallel composition illustrated}
\label{fig:para-illust}
\end{figure}


The example $A \parallel B$ of untimed I/O transition systems\footnote{Convention: plain states are unmarked while the $\top$ and $\bot$ states are marked by $\top$ and $\bot$ resp. To simplify drawing, multiple copies of $\top$ and $\bot$ are allowed but the self-loops on them are omitted.} in Figure~\ref{fig:para-illust} shows the case of parallel composition of two processes, one with output $a$ disabled and the other with input $a$ enabled. According to classic semantics, this will produce an output which is disabled. If we move the example into the level of $\top$/$\bot$-completed TIOTSs (i.e.  $A' \parallel B'$), this means $\top$ in parallel with a plain state gives rise to the product state $\top$ (i.e. $\top \parallel p = \top$). Similarly, if we have two processes $C \parallel D$, on which input $a$ is disabled on one process and output $a$ is enabled on the other, then their parallel composition should generate an output action leading to $err$, which if mapped into the level of $\top$/$\bot$-completed TIOTSs gives rise to $\bot \parallel p = err$. The $err$ state models \emph{error-trapping} states like those employed in the mechanisms of exception or timeout. Since we cannot interpret $err$ as $\top$, the only option left is to interpret it as $\bot$. This gives rise to our definition of the parallel composition.

\paragraph{Parallel composition} 
Starting with the parallel composition operator, this paper will introduce a series of four operators for process composition, all of which are a variant of the synchronised product operator. In order to obtain a modular structure and factor out the variations amongst operators, we adopt a two-step approach. In the first step we define a state composition operator and an alphabet composition operator. In the second step, we use the \emph{state-to-process lifting technique}, defined as a generic synchronised product operator, to lift the composition to the process level. 

A \emph{generic synchronised product} operation $\prod_{\otimes}$ is a binary process composition operation parameterised by another binary \emph{polymorphic} operation $\otimes$. That is, $\otimes$ needs to be defined both as a \emph{state composition} operation and as an \emph{alphabet composition} operation. 

\paragraph{State-to-process lifting} Given two $\top/\bot$-completed TIOTS, $\P_i= \langle I_i, O_i, S_i, s_i^0, \rightarrow_i \rangle$ for $i \in \{0,1\}$, satisfying $S_0\cap S_1 = \{\bot,\top\}$, $\P_0 \prod_{\otimes} \P_1$ gives rise to a new $\top/\bot$-completed TIOTS $P= \langle I, O, S, s^0, \rightarrow\rangle$ s.t. $(I,O) = (I_0, O_0) \otimes (I_1, O_1)$, $S = (P_0 \times P_1) \uplus P_0 \uplus P_1 \uplus \{\top, \bot\}$, $s^0 = s^0_0 \otimes s^0_1$ and $\rightarrow$ is the smallest relation containing $\rightarrow_0 \cup \rightarrow_1$,\footnote{Containment of $\rightarrow_0 \cup \rightarrow_1$ is not required for parallel composition definitions but is so for conjunction and disjunction definitions in the sequel.} and satisfying the rules:
\begin{center}
\Large
\noindent
${p_0 \ar{\alpha}_{0} s_0'} \ \ {p_1 \ar{\alpha}_{1} s_1'} \over {p_0 \otimes p_1 \ar{\alpha} s_0' \otimes s_1'}$  \ \
\vspace{2mm} ${p_0  \ar{a}_0 s_0'} \ \  a \notin A_1  \over {p_0 \otimes p_1} \ar{a} {s_0' \otimes p_1}$ \ \
\vspace{2mm} ${p_1  \ar{a}_1 s_1'} \ \  a \notin A_0  \over {p_0 \otimes p_1} \ar{a} {p_0 \otimes s_1'}$ \ \
\normalsize
\end{center}

The parallel composition operation is an instantiation of the generic synchronised product by the polymorphic operation $\parallel$, i.e. $\prod_{\parallel}$. The associated interpretation of $s_0 \parallel s_1$ is supplied in Table~\ref{table:parallel} while $(I_0, O_0) \parallel (I_1, O_1)$ is defined to be $((I_0 \cup I_1) \setminus (O_0\cup O_1), O_0\cup O_1)$ under the assumption that $O_0 \cap O_1 = \{\}$, i.e. $\P_0$ and $\P_1$ have \emph{$\parallel$-composable alphabets}.

In Table~\ref{table:parallel} the $\parallel$-product state is in $\top$ (or $\bot$) if one of the component states is in $\top$ (or $\bot$). If they are simultaneously (i.e. one each) in $\top$ and $\bot$, $\top$ will have priority and the product will be $\top$.\footnote{If the TIOTSs are derived from TIOAs with disjoint clocks, then we define $p_0 \times p_1$ for plain states $p_i=(l_i, t_i)$ with $i \in \{0,1\}$ as $((l_0,l_1),t_0 \uplus t_1)$.}

\begin{table}[t]
\caption{State $||$-product.}
\begin{tabular}{l | l c l}
$\boldsymbol\parallel$ &  \ \ $\top$ & $p_0$ &  \ \ $\bot$ \\
\hline
$\top$ &  \ \ $\top$ \ \ & \ \ $\top$ \ \ & \ \ $\top$ \ \ \ \  \\
$p_1$ &  \ \ $\top$ \ \  & \ \ $p_0 \!\!\cross\!\! p_1$ \ \ & \ \ $\bot$ \\
$\bot$ & \ \  $\top$ \ \ & \ \ $\bot$ \ \ & \ \ $\bot$
\end{tabular}
\label{table:parallel}
\end{table}

The definition of the parallel operator can be lifted to TIOAs (c.f. \ref{app:a}). 

\subsection{Incompatibility errors and timelocks}

When two components are composed, the parallel composition automatically checks whether the guarantees provided by one component meet the assumptions required by the other. For instance, the arrival of an input at a location and time of a component when it is not expected (i.e. the input is disabled at the location and time) triggers a \emph{safety error} (aka exception) in the parallel composition. Or the non-arrival of an expected input at a location before its timeout (specified by the co-invariant) triggers a \emph{bounded-liveness error} (aka timeout) in the parallel composition.

Formally, we have two possible ways to characterise the incompatibility errors (i.e. exception and timeout), one based on closed systems while the other on open systems. 

For closed systems, it is obvious that safety errors are simply actions (i.e. output) transitions leading to $\bot$, while bounded-liveness errors are delay transitions leading to $\bot$. Thus a closed system is free of incompatibility errors iff it is free of $\bot$, i.e. $\bot$ is not reachable in the system.  This characterisation is very robust, working for both the theory with the timestop capability and the theory without. Actually, we will use it as a basis for defining the refinement relations in both theories. The first refinement will be used as an stepping stone to build the second one.


For open systems, however, the characterisation is less obvious. Below we use detailed analysis of two examples to illustrate incompatibility errors. Note that the open-system characterisation is only meaningful for the theory without the capability to stop time. For the theory with timestop capability, since an environment can use $\top$ to steer any component out of $\bot$, it is not meaningful to examine incompatibility errors before a system is fully closed.

\begin{figure}[t]
\begin{center}
\includegraphics[width=0.55\textwidth]{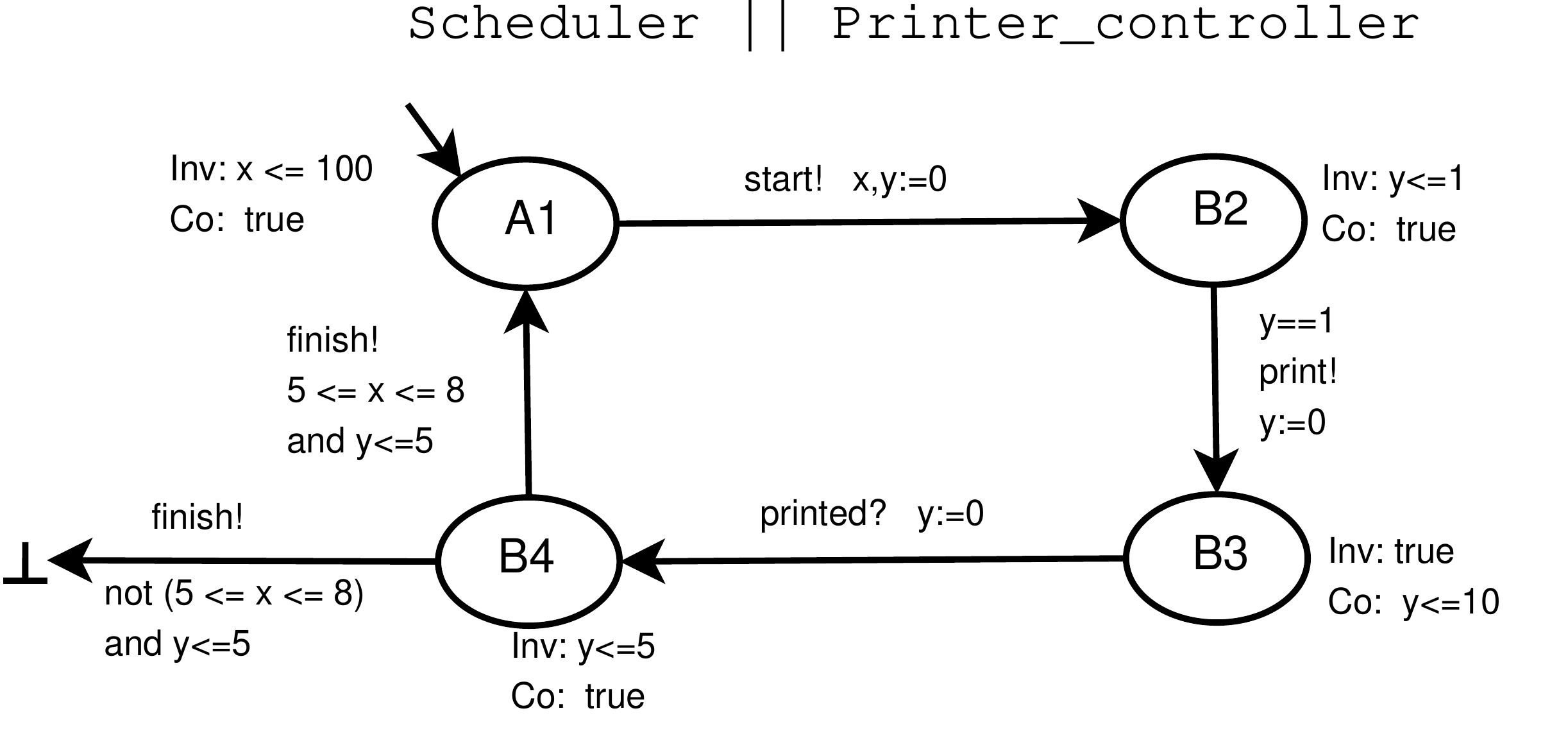}
\end{center}
\caption{Parallel composition of the job scheduler and printer controller.}
\label{fig:product}
\end{figure}

\paragraph{Examples: exception} Figure~\ref{fig:product} shows the parallel composition of the job scheduler with the printer controller (c.f. \ref{app:a}). In the transition from $B4$ to $A1$, the guard combines the effects of the constraints on the clocks $x$ and $y$. As $finish$ is an output of the controller, it can be fired at a time when the scheduler is not expecting it, meaning that an exception is raised due to safety errors. This is indicated by the transition to $\bot$ when the guard constraint $5\leq x \leq 8$ is not satisfied.

Technically speaking, an exception is modelled by auto-$\bot$. We say a plain state $p$ is an \emph{auto-$\bot$ state} iff $p \ar{a} \bot$ for some $a \in O$. Obviously auto-$\bot$ is insensitive to $\bot$-removal.

Intuitively, an exception is an \emph{uncontrollable} (i.e. by the environment) action transition to $\bot$, i.e. the system can independently execute the action transition and go to $\bot$  no matter how the environment behaves. In contrast, a TIOTS might also have \emph{controllable} action transitions to $\bot$, e.g. input transition to $\bot$, whose occurrence depends more on the environment than the system. 


\begin{figure}[t]
\begin{center}
\includegraphics[width=\textwidth]{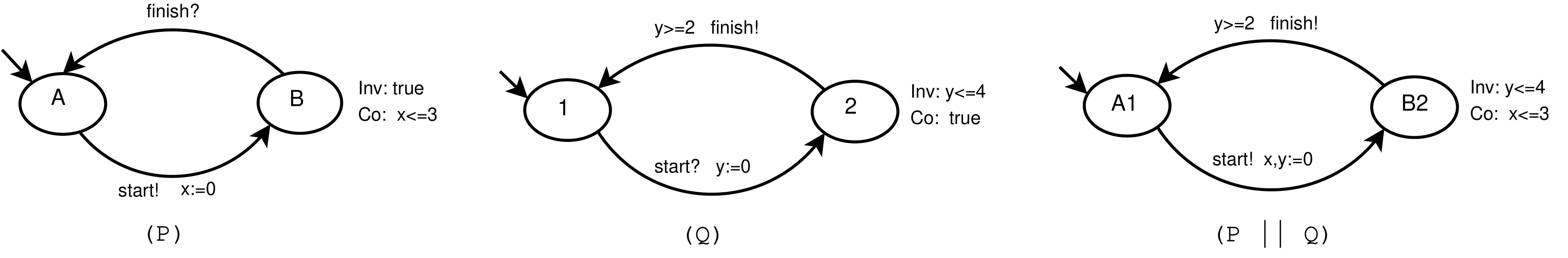}
\end{center}
\caption{Bounded liveness error.}
\label{fig:BLerr}
\end{figure}

\vspace{3mm}

\paragraph{Examples: timeout}

Another example to show bounded-liveness errors is given in Figure~\ref{fig:BLerr}. In the closed system $\P \parallel \Q$, at location $B2$ the system is free to choose either output $finish$ after $y \geq 2$ or delay until $x>3$. If it chooses the latter, $\P$ component will time out in location $B$ and the system will enter $\bot$. Note that the timeout here is due to the fact that the urgency requirement at location $2$ of $\Q$ (i.e. $y <=4$) is weaker than the timeout bound set at location $B$ of $\P$ (i.e. $x <=3$). (If it is otherwise, the invariant at $B2$ will preempt the co-invariant at $B2$ and eliminate the possibility of timeout.)

Technically speaking, a timeout is modelled by semi-$\bot$. We say a plain state $p$ is a \emph{semi-$\bot$ state} iff 1) all input transitions in $p$ or any of its time-passing successors lead to $\bot$, and 2) there exists $d \in \mathbb{R}^{>0}$ s.t. $p \ar{d} \bot$. Thus a semi-$\bot$ represents a point in time from which on the environment has no safe input that it can use to interrupt the system's delay process into $\bot$. Our definition is based on $\top/\bot$-complete TIOTSs. It is easy to see semi-$\bot$ is not affected by $\bot$-removal. Thus we can extrapolate the definition onto $\top/\bot$-removed TIOTSs as well.

Intuitively, a timeout is an \emph{uncontrollable} delay transition to $\bot$, i.e. the system can independently execute the delay transition and go to $\bot$ no matter how the environment behaves. In contrast, a TIOTS might also have \emph{controllable} delay transitions to $\bot$, e.g. delay transition to $\bot$ with input exits, where the environment can interrupt the delay process by inputting at the proper moment. In Section~\ref{sec:realise} we will use timed games to formalise these intuitions. 

For open systems, a $\bot$-free TIOTS is free of auto-$\bot$ but is not necessarily free of semi-$\bot$. Indeed, $\bot$-freedom here is neither a sufficient nor necessary condition for an open system to be free of incompatibility errors, which, instead, corresponds (informally) to a system free of auto-$\bot$ and semi-$\bot$. A more formal definition will have to wait until Section~\ref{sec:realise}.





\

Similarly to equating $\bot$ to the error-trapping state of classical I/O systems, we can also explain $\top$ within the classical I/O framework (i.e. without relying on intuitions like assumption/guarantee violations) by augmenting it with a \emph{timestop} state. Timestop models the operation of stopping the system clock and in our context means the freezing of global time. We equate $\top$ to timestop. Thus, $\top$ represents the \emph{magic moment} from which the global time (or the whole system) stops elapsing (or running), consequently eliminating, once and for all, all subsequent possibility of errors. From an environment's point of view we assume that $\top$ refines plain states, which in turn refine $\bot$. Timestop can explain the behaviour of $\top$ in parallel composition: the equation $\bot \parallel \top = \top$ holds because time stops exactly at the moment the error-trapping mechanism is triggered, so the resulting state is a timestop, rather than $\bot$. 

Dual to auto-$\bot$ and semi-$\bot$, we can also define notions like auto-$\top$ and semi-$\top$. We say a plain state $p$ in a $\top/\bot$-complete TIOTS is an \emph{auto-$\top$} iff $p \ar{a} \top$ for some $a \in I$. We say a plain state $p$ is a \emph{semi-$\top$} iff 1) all output transitions in $p$ or any of its time-passing successors lead to the $\top$ state, and 2) there exists $d \in \mathbb{R}^{>0}$ s.t. $p \ar{d} \top$. 

We cannot fully explain the intuitions behind auto-$\top$ at this stage. But, for semi-$\top$, it models a generalisation of timelock to open systems. Here we need to switch back to $\top$-removed semantics for TIOTSs, where the intuition of timelock is clearer.

On a closed ($\top$-removed) TIOTS, the definition of timelock coincides with that on classical TAs\footnote{Due to our non-zenoness assumption, our timelock can be shown to be a local and strengthened version of the timelock defined as in~\cite{Baier08}.} (i.e. TAs without I/O distinction). We call a plain state $p$ a \emph{timelock} if 1) no \emph{action} transition is enabled in $p$ or any of its time-passing successors, and 2) there exists $d \in \mathbb{R}^{>0}$ s.t. $d$ is not enabled in $p$. 

\paragraph{Semi-$\top$ as timelock for open systems} The definition of semi-$\top$ can be specialised for $\top$-removed TIOTSs. We say a plain state $p$ is a semi-$\top$ iff 1) no \emph{output} transition is enabled in $p$ or any of its time-passing successors, and 2) there exists $d \in \mathbb{R}^{>0}$ s.t. $d$ is not enabled in $p$. Obviously semi-$\top$ is a generalisation of timelock to open systems, which models the scenario that the component has no option but to stop the progress of time if the environment does not intervene in time.




Like the case for $\bot$-freedom, a $\top$-free TIOTS is free of auto-$\top$ but is not necessarily free of semi-$\top$. Thus, timelock is independent of timestop, which confer on the component an \emph{implicit capability} to stop time. 




\

Before moving on to the next section, we make a few observations as summary: 

\begin{itemize}

\item We model errors arising from assumption/guarantee mismatches by auto-$\bot$ and semi-$\bot$ states and we model timelock by semi-$\top$.

\item When two components are composed in parallel, new errors will be generated but no new timelock (or auto-$\top$) will be generated. 

\item This non-duality in the effect of parallel composition is largely due to the non-symmetric treatment of input and output in the parallel composition: the synchronisation of an input and an output gives rise to an output. For example, in Figure~\ref{fig:BLerr},  the component $\P$ in location $B$ is not a semi-$\bot$ since it has an outgoing input transition $finish$. But, after parallel composition, the input becomes output and $B2$ contains a semi-$\bot$.



\end{itemize}

\


\section{Timed I/O Games and Refinement}\label{sec:game}

We have used game-based intuitions to introduce $\top$ and $\bot$ as assumption and guarantee violations resp. Now let us elaborate further and formalise the timed-game framework, whereby the \emph{component} and an \emph{environment}, controlling timed outputs and inputs, respectively, 
play a $\top$/$\bot$-reachability game in which the component tries to avoid reaching $\top$, while the environment tries to avoid reaching $\bot$. Previously there have been works on timed game framework~\cite{Cassez05,henzinger-timedia}. But our formulation has important differences (cf the discussion at the end of Section~\ref{sec:op}).

\subsection{Timed I/O Games}\label{sec:subgame}

In our timed I/O game, a TIOTS encodes the set of strategies possible for the component in the game. An \emph{environment} for a TIOTS $\P$ is any TIOTS $\Q$ such that $\P$ and $\Q$ have \emph{complementary} alphabets, meaning $I_\P=O_\Q$ and $O_\P=I_\Q$.  $\Q$ encodes the environmental strategies.

The formal definition of (timed) strategies is given below:

\begin{itemize}
\item
A \emph{strategy} $\G$ is a deterministic tree TIOTS\footnote{We say an acyclic TIOTS is a \emph{tree} if 1) there does not exist a pair of transitions in the form of $p \ar{a} p''$ and $p' \ar{d} p''$, 2) $p \ar{a} p'' \wedge p' \ar{b} p''$ implies $p=p'$ and $a=b$ and 3) $p \ar{d} p'' \wedge p' \ar{d} p''$ implies  $p=p'$.} s.t. each plain state in $\G$ is ready to accept all possible inputs by the environment, but allows a single move (delay or output) by the component. 

That is, the set of enabled timed actions in any state $p$ of $\G$ is $I \uplus mv_{\G}(p)$, where $mv_{\G}(p)$ is the enabled component move, being either $\{a\}$ for some $a \in O$ or a time interval\footnote{Note that all invariants and co-invariants are downward-closed. Thus, a delay move can be represented as a time interval from $0$ to some $d \in \mathbb{R}^{\geq 0}$ or to infinity.}. The time interval here can be either infinite, i.e. $(0,\infty)$, or finite, i.e. $(0,d]$ for some $d \in \mathbb{R}^{>0}$. (Note that $(0,d]$ is the set of all enabled delay at a state. Thus, due to time additivity, $d$ should be the maximal delay allowable by the strategy TIOTS from that state. In another word, the move proposed at the new state after firing $d$ must be an action move, say $a$.\footnote{That is, at each state a strategy proposes either  a $\langle d, a \rangle$ move (for $d \geq 0$) or a $\infty$ move.})

\item
Given TIOTSs $\P$ and $\P'$ with \emph{identical alphabets} (i.e. $O = O'$ and $I = I'$), we say $\P$ is a \emph{partial unfolding}~\cite{wang12}
of $\P'$ if there exists a function $f:S_\P\rightarrow S_{\P'}$ such that 1) $f$ maps $\top$ to $\top$, $\bot$ to $\bot$ and plain states to plain states, and 2) $f(s^0_{\P})= s^0_{\P'}$ and $p \ar{\alpha}_{\P} s \implies f(p) \ar{\alpha}_{\P'} f(s)$.

\item We say a TIOTS $\P$ \emph{contains} a strategy $\G$ if $\G$ is a partial unfolding of $(\P^{\bot})^{\top}$. 

\item
We say a simple-path TIOTS\footnote{We say an acyclic TIOTS is a \emph{simple path} if 1) $p \ar{a} s' \wedge p \ar{\alpha} s''$ implies $s'=s''$ and $a=\alpha$ and 2) $p \ar{d} s' \wedge p \ar{d} s''$ implies  $s'=s''$.} $L$ is a \emph{run} of $\P$ if $L$ is a partial unfolding of $\P$.

\end{itemize}

The set of strategies\footnote{In this paper we use a set of strategies (say $\Gamma$) to mean a set of strategies with identical alphabets.} contained in $\P$ is denoted as the extension $[\P]$. Since it makes little sense to distinguish strategies that are isomorphic, we will freely use strategies to refer to their isomorphism classes and write $\G=\G'$ to mean $\G$ and $\G'$ are isomorphic.

\begin{figure}[t]
\begin{center}
\includegraphics[width=0.8\textwidth]{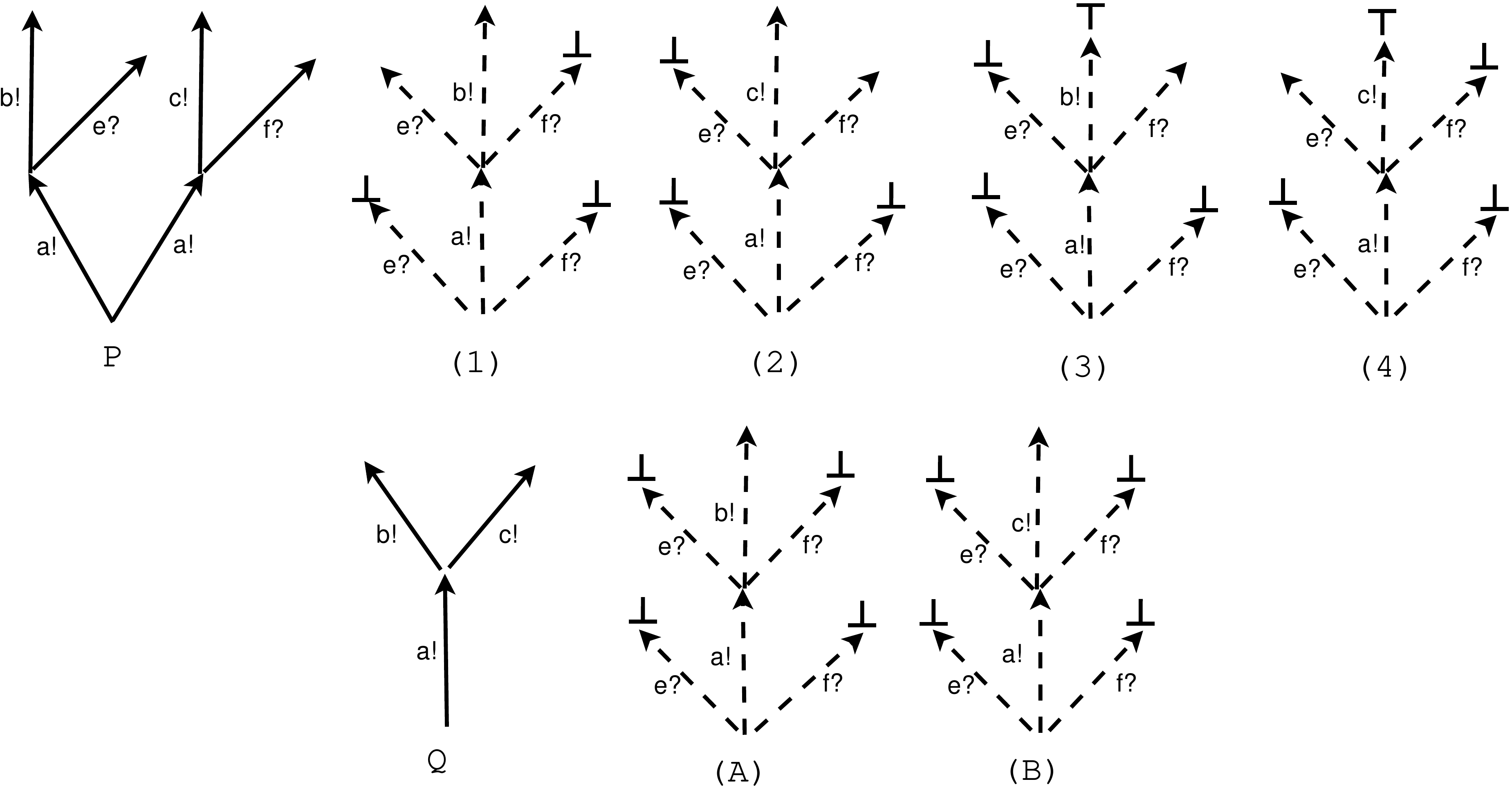}
\end{center}
\caption{Strategy example.}
\label{fig:strategy-equivalence-strong}
\end{figure}

Let us give some examples in Figure~\ref{fig:strategy-equivalence-strong}. For the sake of simplicity we use two untimed transition systems $P$ and $Q$, with identical alphabets $I=\{e,f\}$ and $O=\{a,b,c\}$, to illustrate the idea of strategies. The transition systems use solid lines while strategies use dotted lines. We show four strategies of $P$ and two strategies of $Q$ on the right hand side of $P$ and $Q$ resp. in Figure~\ref{fig:strategy-equivalence-strong}. (They are not the complete sets of strategies for $P$ and $Q$.) Note that the strategies $3$ and $4$ owe their existence to the $\top$-completion.

\paragraph{Game rules} When a component strategy $\G$ is played against an environment strategy $\G'$, at each game state (i.e. a product state $p_{\G} \times p_{\G'}$) $\G$ and $\G'$ each propose a move (i.e. $mv_{\G}(p_{\G})$ and $mv_{\G'}(p_{\G'})$). If one of them is a delay and the other is an action, the action will prevail. If both propose delay moves (i.e. $mv_{\G}(p_{\G}), mv_{\G'}(p_{\G'}) \subseteq \mathbb{R}^{\geq 0}$), the smaller one (w.r.t. set containment) will prevail.

Since a delay move proposed at a strategy state is the maximal delay allowable at that state and the next move must be an action move, a play cannot have two consecutive delay moves.

If, however, both propose action moves, there will be a tie, which will be resolved by tossing the coin. For uniformity's sake, the coin can be treated as a special component. A strategy of the coin is a function $h$ from $tA^*$ to $\{0,1\}$. We denote the set of all possible coin strategies as $H$.

\paragraph{Remark} Our game rules are consistent with those found in~\cite{henzinger-timedia,Asarin98}. But our use of the rules is different. In~\cite{henzinger-timedia,Asarin98}, there is no restriction that the rules must be applied on a pair of pre-determined strategies that propose only maximal delay moves. So if both players propose delay moves in one round, the winning side (with smaller delay) can still propose a second delay move in the next round. This creates complications like time-blocking strategies and blame assignment~\cite{henzinger-timedia}.


\paragraph{Strategy composition} A play of the game can be formalised as a composition of three strategies, one each from the component, environment and coin, denoted $\G_{\P} \cross_h \G_{\Q}$. At a current game state $p_{\P} \times p_{\Q}$, if the prevailing action is $\alpha$ and we have $p_{\P} \ar{\alpha} s_{\P}$ and $p_{\Q} \ar{\alpha} s_{\Q}$, then the next game state is $s_{\P} \parallel s_{\Q}$. The play will stop when it reaches either $\top$ or $\bot$. The composition will produce a simple path $L$ that is a run of $\P \parallel \Q$,\footnote{$\P \parallel \Q$ gives rise to a \emph{closed system} (i.e. the input alphabet is empty), a run of $\P \parallel \Q$ is a strategy of $\P \parallel \Q$.} i.e. either an infinite plain run or a finite run ending in $\top$/$\bot$. There is no possibility of finite plain run, as is possible in~\cite{henzinger-timedia,Asarin98} by playing an infinite sequence of delay moves that converges. 

Strategy composition can be generalised to composition between any pair of strategies $\G_{\P} \cross_h \G_{\Q}$ with \emph{$\parallel$-composable alphabets}. That is, $O_{\P} \cap O_{\Q} = \{\}$. For such $\P$ and $\Q$, $\G_{\P} \cross_h \G_{\Q}$ gives rise to a tree rather than a simple-path TIOTS. That is, at each game state $p_{\P} \times p_{\Q}$, besides firing the prevailing $\alpha \in tO_{\P} \cup tO_{\Q}$, we need also to fire 1) all the synchronised inputs, i.e. $e \in I_{\P} \cap I_{\Q}$, and reach the new game state $s_{\P} \parallel s_{\Q}$ (assuming $p_{\P} \ar{e} s_{\P}$ and $p_{\Q} \ar{e} s_{\Q}$) and 2) all the independent inputs, i.e. $e \in (I_{\P} \cup I_{\Q}) \setminus (A_{\P} \cap A_{\Q})$, and reach the new game state $s_{\P} \parallel p_{\Q}$ or $p_{\P} \parallel s_{\Q}$.

The generalisation enables us to reduce parallel composition on processes to strategy composition:


\begin{lemma}
For $\parallel$-composable TIOTSs $\P$ and $\Q$, $[\P \parallel \Q] = [\P] \cross [\Q]$, where we define $\Gamma \cross \Gamma' = \{\G \cross_h \G' \, | \, \G \in \Gamma, \G' \in \Gamma'$ and $h \in H\}$.



\end{lemma}

\subsection{Refinement, Determinisation and Strategy Characterisation}\label{sec:ref-det}

A TIOTS is a refinement of another if it will work in any environment that the original worked in without introducing safety or bounded-liveness errors. Here we use the \emph{the closed system version} of incompatibility errors to formulate the definition. 

\begin{definition}[Substitutive Refinement]
\label{defn:op-refine}
Let $\P_{imp}$ and $\P_{spec}$ be TIOTSs with identical alphabets. $\P_{imp}$ \emph{refines} $\P_{spec}$, denoted $\P_{spec} \sqsubseteq \P_{imp}$, iff for all environments $\Q$, $\P_{spec} \parallel \Q$ is $\bot$-free implies $\P_{imp} \parallel \Q$ is $\bot$-free. We say $\P_{imp}$ and $\P_{spec}$ are \emph{substitutively equivalent}, i.e. $\P_{spec} \simeq \P_{imp}$, iff $\P_{imp} \sqsubseteq \P_{spec}$ and $\P_{spec} \sqsubseteq \P_{imp}$.
\end{definition}


Alternatively, if we view $\P_{imp}$ and $\P_{spec}$ as two $\bot$-reachability games and replace parallel composition by strategy composition, the refinement can be defined as a comparison on how challenging each game is for the environment.
In the games, \emph{the component and coin collaborate trying to reach $\bot$ whilst the environment tries to avoid reaching $\bot$}. Therefore, $\P_{spec} \sqsubseteq \P_{imp}$ iff, all environment strategies winning in game $\P_{spec}$ are also winning in game $\P_{imp}$. Here we say an environment strategy $\G_E$ is \emph{winning} in game $\P$ (or winning against strategy set $[\P]$) iff $\G_E \cross_h \G$ is $\bot$-free for all $\G \in [\P]$ and $h \in H$.


Obviously, $\P \simeq \Q$ is related but not equivalent to the set containment between $[\P]$ and $[\Q]$; $[\Q] \subseteq [\P]$ implies $\P \simeq \Q$ but the converse is not true. This failure of the equivalence is largely due to the phenomenon of implicit strategies. 



Formally, we say a strategy $\G \notin [\P]$ is an \emph{implicit strategy} of $[\P]$ iff all environment strategy winning against strategy set $[\P]$ are also winning against $[\P] \cup \{\G\}$. Thus, a general principle to formulate a strategy-based semantics is to perform some closure operation on $[\P]$ s.t. all implicit strategies become included. 

Given $[\P]$, the set of its implicit strategies depends on the refinement order under consideration. With respect to $\simeq$ there are two sources of implicit strategies. 

The first is due to the existence of an ordering on strategies; some strategies are by nature more aggressive than the others.

\paragraph{Comparing strategies} When the game is played, the component tries to avoid reaching $\top$ while the environment tries to avoid reaching $\bot$. Different strategies in $[\P]$ vary in their effectiveness to achieve the objective. Such effectiveness can be compared if two strategies closely resemble each other: we say $\G$ and $\G'$ are \emph{affine} if $s^0_{\G} \dar{tt} p$ and $s^0_{\G'} \dar{tt} p'$ implies $mv_{\G}(p) = mv_{\G'}(p')$. Intuitively, this means $\G$ and $\G'$ propose the same move at the `same' states. For instance, the strategies $1$, $3$ and $A$ in Figure~\ref{fig:strategy-equivalence-strong} are pairwise affine, and so are the strategies $2$, $4$ and $B$.

Given two affine strategies $\G$ and $\G'$, we say $\G$ is \emph{more aggressive} than $\G'$, denoted $\G \preceq \G'$, if 1) $s^0_{\G'} \dar{tt} \bot$ implies there is a prefix $tt_0$ of $tt$ s.t. $s^0_{\G} \dar{tt_0} \bot$ and 2) $s^0_{\G} \dar{tt} \top$ implies there is a prefix $tt_0$ of $tt$ s.t. $s^0_{\G'} \dar{tt_0} \top$. Intuitively, it means $\G$ can reach $\bot$ faster but $\top$ slower than $\G'$.  $\preceq$ forms a partial order over $[\P]$, or, more generally, over any set of strategies with identical alphabets. For instance, strategy $A$ is more aggressive than $1$ and $3$, while strategy $B$ is more aggressive than $2$ and $4$.

When the game is played, the component $\P$ prefers to use the maximally aggressive strategies in $[\P]$\footnote{This is because our semantics/refinement is designed to preserve $\bot$ rather than $\top$.}. Thus, two components that differ only in non-maximally aggressive strategies should be equated. We define the \emph{strategy semantics} of component $\P$ to be $\semb{\P}=[\P]^{\preceq}$, i.e. the upward-closure of $[\P]$ w.r.t. $\preceq$.

The other source of implicit strategies is due to the imperfect information of our game. That is, given a partial play $tt$ of a non-deterministic game $\P$, there are a number of possible states (say $S_{tt}$) that can be reached. It is the component and coin, not the environment, that knows which of $S_{tt}$ is chosen as the next game state. This entitles the former to have implicit strategies, which are hybrid strategies generated through decomposing and re-combining the strategies of different states in $S_{tt}$. For instance, strategy $A$ is a hybrid of strategies 1 and 3 in Figure~\ref{fig:strategy-equivalence-strong}.

Such implicit strategy can be made explicit by converting an imperfect information game into an (equivalent) perfect information game. Below we propose a modified subset construction procedure to perform such conversion. 




We define the \emph{determinisation} $\P^D$ of a $\bot$-complete TIOTS $\P$ as a modified subset construction procedure on $\P$: given a subset $S_0$ of states reachable by a given trace, we only keep those which are minimal w.r.t. the state refinement relation. So if the current state subset $S_0$ contains $\bot$, the procedure reduces $S_0$ to $\bot$; if $\bot \notin S_0 \neq \{\top\}$, it reduces $S_0$ by removing any possible $\top$ in $S_0$.\footnote{For a more detailed definition of transforming non-deterministic systems into substitutivity-equivalent deterministic systems, we refer readers to the Definition 4.2 in~\cite{WangKwiat07}. That is for the untimed case.} For example, Figure~\ref{fig:strategy-equivalence-strong} contains two $\top/\bot$-removed TIOTSs $P$ and $Q$. If we apply the above procedure to $P^{\bot}$ the resultant TIOTS will be $Q^{\bot}$.

Given any TIOTS $\P$, we can verify $\P \simeq \P^D$ even though $[\P]^{\preceq} \subseteq [\P^D]^{\preceq}$. 

\begin{proposition}[\cite{CKW12}]
\label{prop:1}
Any TIOTS $\P$ is substitutively equivalent to the deterministic TIOTS $\P^D$.
\end{proposition}


For instance, in Figure~\ref{fig:strategy-equivalence-strong} we have $(P^{\bot})^D = Q^{\bot}$, but $[P]^{\preceq} \neq [Q]^{\preceq}$ since $1$, $2$, $3$ and $4$ are strategies of $[Q]^{\preceq}$ (due to upward-closure w.r.t. $\preceq$) but $A$ and $B$ are not strategies of $[P]^{\preceq}$.

There might be further sources of implicit strategies with respect to coarser refinements than $\simeq$. But, for the two sources of $\simeq$, we can give a uniform and collective characterisation. That is, we say a strategy $\G' \notin \Gamma$ is a \emph{$\simeq$-implicit strategy} of the strategy set $\Gamma$ iff $s^0_{\G'} \dar{tt} s'$ implies there exists $s^0_{\G} \dar{tt} s$ for some $\G \in \Gamma$ s.t. either both executions are plain executions or execution $s^0_{\G'} \dar{tt} s'$ reaches $\top$ earlier or $\bot$ later than $s^0_{\G} \dar{tt} s$. We denote by $\Gamma^E$ the \emph{$\simeq$-implicit strategy closure} of $\Gamma$.

Define $\semb{\P} = [\P]^E$. Then $\semb{\cdot}$ characterises exactly the substitutive equivalence $\simeq$.

\begin{theorem}[\cite{CKW12}]
\label{thm:1}
Given TIOTSs $\P$ and $\Q$,  $\P \sqsubseteq \Q$ iff $\semb{\Q} \subseteq \semb{\P}$.
\end{theorem}




\section{Realisability Restriction and Coarsened Refinement}
\label{sec:realise}

Section~\ref{sec:ref-det} gives a substitutive refinement and its strategy characterisation. \cite{CKW12} further prove that $\simeq$ is a congruence w.r.t. the parallel, conjunction, disjunction and quotient operators, thus giving rise to a simple and elegant compositional specification theory.\footnote{Actually the theory in~\cite{CKW12} is developed in a more general setting, where the assumption of non-zenoness is removed.}

However, one drawback of such a theory is that we allow \emph{unrestricted strategies} for the component and environment in the game play. In another word, the component and environment may apply timestop-like operations (i.e. timestop and timelock) directly against each other. 

The timestop-like operations greatly increase the distinguishing power of the environment, giving rise a finest possible equivalence $\simeq$.  It also equips the environment with the capability to steer components away from incompatibility errors ($\bot$) under all possible situations, thus making conjunction and quotient a fully defined operator. 

In general, such capability is too powerful to be realistic.
Certain real-world systems might have an inherent ability to stop the system clock, e.g. in embedded systems and circuit design~\cite{Lim86,MTCMR00} or in a \emph{controlled execution environment} like simulation or testing. However, for even larger class of applications, the suspension of clocks is arguably neither meaningful nor realisable.

Thus, in the rest of the paper we will develop a theory that can remove timestops and timelocks, to keep only the so-called realisable behaviours. Note that, even for such timestop-free systems, $\top$ can play the important role of being an imaginary state exploited at the intermediate steps of theory development and thus greatly simplifying operator definitions like quotient and conjunction.

 



We focus on realisable systems from hereon, and simply call TIOTSs free of $\top$ and semi-$\top$\footnote{This, combined with our non-zenoness assumption on TIOTSs, implies that no component in our realisable theory is time-blocking.}  \emph{specifications}. Therefore, we are returning to the classical I/O systems equipped with error-trapping states. As can be demonstrated, operations on components such as parallel composition, renaming, hiding and determinisation preserve $\top$ and semi-$\top$ freedom\footnote{This is in contrast to the case of synchronised product on timed components without I/O distinction, where new timelocks can be generated.}.

Hence, we offer a classical I/O system as a user interface so that complications like timestops and timelocks are hidden from view and components and environments use only realisable strategies to interact with one another. Formally we say a strategy is \emph{realisable} iff it is free of $\top$ and semi-$\top$. 
We often use $\R$ to denote a realisable strategy.




The rest of this section leaves the world of $\top/\bot$-complete TIOTSs and deals exclusively with specifications. Furthermore, we assume all specifications are $\bot$-complete in order to simplify presentation. 

The definition of $\prod_{\parallel}$ (and hence $\parallel$) can be extended without modification to work on $\bot$-complete TIOTSs.\footnote{\label{fn:vee}With the extension, synchronisation failures, i.e. an action being enabled on one process but not so on the other, becomes possible.} As parallel composition preserves $\top$ and semi-$\top$ freedom, $\parallel$ can be directly used as an operation on specifications. In addition, since strategies are $\bot$-complete TIOTSs, we can freely parallel-compose a strategy with a component in the sequel.

\paragraph{Realisable refinement} Based on the parallel operator we can re-define the substitutive refinement on top of specifications: Let $\P$ and $\Q$ be specifications with identical alphabets. $\P$ \emph{realisably refines} $\Q$, denoted $\Q \sqsubseteq_r \P$, iff, for all environment specifications $\N$, $\Q \parallel \N$ is $\bot$-free implies $\P \parallel \N$ is $\bot$-free. We say $\P$ and $\Q$ are \emph{substitutively equivalent}, i.e. $\Q \simeq_r \P$, iff $\P \sqsubseteq_r \Q$ and $\Q \sqsubseteq_r \P$.


Note that in the definition 1) both the component and environment are restricted to realisable ones and 2) the incompatibility errors utilised are the closed system version. It is obvious that $\simeq_r$ is the weakest equivalence preserving $\bot$. In the sequel we show that $\simeq_r$ is a congruence w.r.t. the parallel $\parallel$, conjunction $\wedge$, disjunction $\vee$ and quotient $\%$ operators.

Recall that our determinisation is directly defined on $\bot$-complete TIOTSs. On specifications, it is easy to verify that determinisation preserves $\top$ and semi-$\top$ freedom as well as the substitutive equivalence, i.e. $\P \simeq_r \P^D$.

With determinisation, imperfect-information games can be converted into perfect-information games. Based on the latter, we can formalise the notion of incompatibility errors for open systems.

Given a perfect-information game $\P^D$ in which the collaboration of the component and coin play against the environment for the objective of $\bot$-reachability, we say a plain state $p$ in $\P^D$ is \emph{$\bot$-winning} iff there is no (realisable) environment strategy winning in game $\P^D(p)$. In another word, starting from state $p$, the component and coin can collaborate to win the $\bot$-reachability game. Here we use the notation $\P(p)$ to denote the specification $\P$ with the initial state changed to $p$. 

Obviously, semi-$\bot$ and auto-$\bot$ states are $\bot$-winning states (under realisability restriction) and without realisability restriction no state in game $\P^D$ is $\bot$-winning. 

Semi-$\bot$ and auto-$\bot$ are one of the most representative subclass of $\bot$-winning states; the absence of semi-$\bot$ and auto-$\bot$ effectively captures the absence of $\bot$-winning states. 

\begin{lemma}
\label{lem:w2as}
A deterministic specification is free of $\bot$-winning states iff it is free of semi-$\bot$ and auto-$\bot$.
\end{lemma}


Based on this observation we can formalise the notion of incompatibility error freedom for open systems. We say an (open) TIOTS $\P$ is \emph{error-free} iff $\P^D$ is free of auto-$\bot$ and semi-$\bot$. From this definition it is easy to see that the perfect information requirement is necessary here since determinisation can introduce new semi-$\bot$. 





\subsection{Strategy characterisation of $\simeq_r$}

The definition of strategies and notation $[\P]$ can be reused on specifications. It is easy to verify that specifications contain only realisable strategies and specification $\parallel$-composition can be reduced to (realisable) strategy composition: $[\P \parallel \Q] = \{\R \cross_h \R' \, | \, \R \in [\P], \R' \in [\Q]$ and $h \in H\}$ for all specifications $\P$ and $\Q$.  

Similarly, we can compare realisable strategies and define $\preceq_r$ as a restriction of $\preceq$ to realisable strategies. This gives rise to the implicit strategy closure operation $\Gamma^{E_r}$ and we define $\semb{\P}_r=[\P]^{E_r}$.









It is easy to verify $\semb{\P}_r = \semb{\Q}_r$ implies $\P \simeq_r \Q$, but the converse is not true. Thus, $\simeq_r$ is strictly coarser than $\semb{\cdot}_r$. 

\begin{figure}
\includegraphics[scale=0.4]{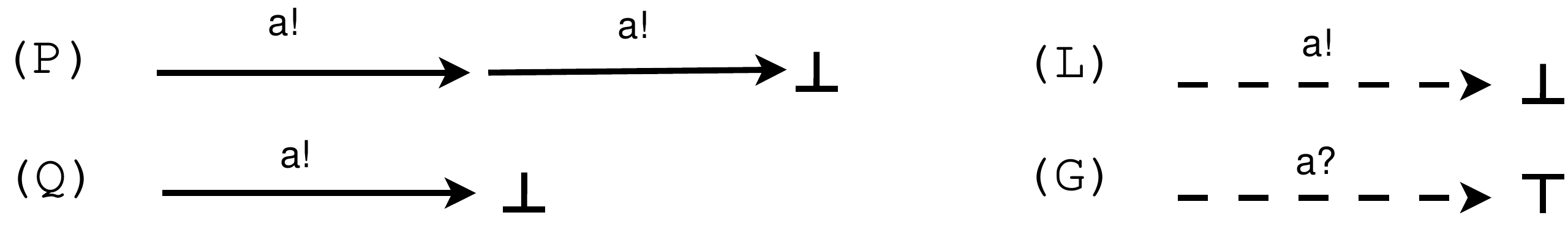}
\caption{Distinguishing power of $\top$.}
\label{fig:distinct1}
\end{figure}

\paragraph{Example} In Figure~\ref{fig:distinct1}, assuming the alphabet is $A= \{a\}$, we were able to distinguish $P$ from $Q$ using $\semb{\cdot}_r$, since strategy $L$ is in $\semb{Q}_r$ but not in $\semb{P}_r$. On the other hand, $P \simeq_r Q$ holds since it is impossible to construct an environment specification $\N$ s.t.  $P \parallel \N$ is $\top$-free but $Q \parallel \N$ is not.




The substitutive equivalence is due to the fact that the initial states of $P$ and $Q$ are both $\bot$-winning states. A $\bot$-winning state is as bad as the $\bot$ state since, once a specification reaches $\bot$-winning states, no (realisable) environment can steer it away from $\bot$. Thus, according to $\preceq_r$ a component in $\bot$-winning states is indistinguishable to one in the $\bot$ state.\footnote{This is in contrast to unrealisable systems, where the environment can always distinguish the $\bot$ state from the $\bot$-winning states by stopping time immediately. For example, the unrealisable strategy $G$ in Figure~\ref{fig:distinct1} can distinguish $P$ from $Q$.} This gives rise to the \emph{third source of implicit strategies}, e.g. strategy $L$ is an implicit strategy of $Q$.





We can make such implicit strategies explicit by performing a further \emph{normalisation} on $\P$.

\paragraph{Normalisation} The normalisation of a specification $\P$, denoted $\P^N$, is obtained by first determinising $\P$ and then collapsing all $\bot$-winning states in $\P^D$ to $\bot$.


An interesting observation here is that normalisation based on $\bot$-winning states can be reduced to normalisation based on semi-$\bot$ and auto-$\bot$, since the latter are those $\bot$-winning states which are precisely one-step away from $\bot$. So we have an alternative \emph{local characterisation} of normalisation. 

$\P^N$ may then be defined by \emph{$\bot$-backpropagation}, which repeatedly collapses semi-$\bot$ and auto-$\bot$ states in $\P^D$ to $\bot$, until semi-$\bot$ and auto-$\bot$ freedom is obtained. 


Since realisable strategies are specifications, normalisation is also defined on realisable strategies. 

\begin{lemma}
Given any component strategy $\R$ and environment specification $\N$, $\R \parallel \N$ is $\bot$-free iff $\R^N \parallel \N$ is $\bot$-free.
\end{lemma}

The normalisation of a specification can be reduced to strategy normalisation. For a set of realisable strategies $\Gamma$, the \emph{normalisation closure}, denoted $\Gamma^{N}$, is the least $\preceq_r$-upward closed superset of $\Gamma$ such that $\R \in \Gamma^N$ implies $\R^N \in \Gamma^N$\footnote{The semantics normalisation operation preserves the disjunction closedness.}.  

\begin{lemma}
Given any specification $\P$, $\semb{\P}_r=\Gamma$ implies $\semb{\P^N}_r = \Gamma^N$.
\end{lemma}


As a shorthand, we use $\semb{\P}_n$ to denote $(\semb{\P}_r)^N$ or $\semb{\P^N}_r$.



\begin{theorem}
\label{thm:3}
Given two specifications $\P$ and $\Q$, $\P \sqsubseteq_r \Q$ iff $\semb{\Q}_n \subseteq \semb{\P}_n$.
\end{theorem}

A specification $\P$ is \emph{inconsistent} iff $s^0_{\P}$ is a $\bot$-winning state. Under normalisation, any inconsistent specification is reduced to the $\bot$-TIOTS. For consistent specifications, normalisation yields a deterministic error-free specification. 








\subsection{Desiderata of the operators}

Before developing the operational definitions on conjunction, disjunction and quotient, let us first describe the desired effects for these operators to achieve.

We say a set of realisable strategies $\Gamma$ is a \emph{specification semantics} iff $\Gamma = (\Gamma^{E_r})^N$. The domain of specification semantics combined with the $\subseteq$ relation gives rise to a lattice, where conjunction ($\wedge$) and disjunction ($\vee$) are supposed to correspond to the join and meet operators respectively.\footnote{As we write $A \sqsubseteq B$ to mean $A$ is refined by $B$, our operators $\wedge$ and $\vee$ are reversed in comparison to the standard symbols for meet and join.} That is, conjunction yields the coarsest specification that is a refinement of its operands, while disjunction yields the finest specification that is refined by both of its operands.


\begin{definition}
For any pair of specification semantics $\Gamma$ and $\Gamma'$ with identical alphabets, we define $\Gamma \wedge \Gamma' = \Gamma \cap \Gamma'$ and $\Gamma \vee \Gamma' = ((\Gamma \cup \Gamma')^{E_r})^N$.
\end{definition}

\noindent
It is easy to verify $\Gamma \cap \Gamma'$ is a specification semantics.

Quotient $\P_0 \% \P_1$ produces the coarsest specification $\P$ such that $\P \parallel \P_1$ is a refinement of $\P_0$. In other words, if $\P_1$ is the \emph{plant} and $\P_0$ is the overall system specification, then $\P_0 \% \P_1$ synthesise the coarsest (or most permissive) \emph{controller} that can steer the plant away from behaviours violating $\P_0$.

Mirror $\P^{\neg}$ gives the set of (realisable) environment strategies that can steer $\P$ away from $\bot$.


\begin{definition}
Given a specification semantics $\Gamma$, we define $\Gamma^{\neg} = \{\R_{\neg}  \, | \, \forall \R \in \Gamma, h \in H: \R \cross_h \R_{\neg} $\ is $\bot$-free$ \}$. Given two specification semantics $\Gamma$ and $\Gamma'$ (with alphabets $A' \subseteq A$ and $O' \subseteq O$), we define $\Gamma \% \Gamma' = \{\R_{\%}  \, | \, \forall \R' \in \Gamma', h \in H: \R_{\%} \cross_h \R' \in  \Gamma \}$.
\end{definition}

\noindent
It is easy to verify $\Gamma^{\neg}$ and $\Gamma \% \Gamma'$ as defined above give rise to specification semantics.



\section{Operational semantics}
\label{sec:opsem}

In the last section we outlined the desiderata for the four operators. Conjunction and disjunction calculate the meet and join w.r.t. $\preceq_r$, whilst mirror and quotient synthesise realisable controllers to steer components away from undesirable states/behaviours. In this section, we give the operational definitions to the operators that fulfill the desiderata. The key challenge here lies in understanding the interplay between synthesis games across specification boundary. 


We adopt a two-step approach here. Firstly we define the four operators for the restricted case when the operands are all normalised specifications. Since the synthesis game in a normalised specification has been pre-resolved, the operator definitions need only to utilise the process-algebraic technique of state-to-process lifting. The process-algebraic definitions may, however, generate a new realistion game under some operators, which, we show, is resolvable by a $\top$-backpropagation procedure.


Then we analyse and understand the composability of different games under different operators; and based on the knowledge we give the minimal extension to the process-algebraic definitions so that the extended operators indeed implement the desiderata for general specifications.

\subsection{Restricted case}
\label{sec:opsem1}


Like parallel composition we define conjunction, disjunction and quotient as variants of synchronised product, which operate over $\top/\bot$-complete TIOTSs and are parameterised by a polymorphic state/alphabet composition operator. 

Table~\ref{table:composition} tells us how states should be combined under the composition operators. Based on the refinement ordering on states, it is easy to see that \emph{state conjunction} ($\wedge$) and \emph{disjunction} ($\vee$) operations in Table~\ref{table:composition} follow the intuition of the join and meet operations (except for the case when both operands are plain states) and that the \emph{state quotient} ($\%$) operation is definable via the \emph{state parallel} ($\parallel$) and \emph{mirror} ($\neg$) operations: $s_0 / s_1 = (s_0^{\neg} \parallel s_1)^{\neg}$.


\footnotesize
\begin{table}[t]
\caption{State composition operators.}

\begin{minipage}[b]{0.26\linewidth}
\centering
\begin{tabular}{l | l c l}
$\boldsymbol\wedge$ &  $\top$ & $p_0$ & $\bot$ \\
\hline
$\top$ &  $\top$ & $\top$ & $\top$ \\
$p_1$ &  $\top$ & $p_0 \!\!\cross\!\! p_1$ & $p_1$ \\
$\bot$ &  $\top$ & $p_0$ & $\bot$
\end{tabular}
\end{minipage}
\hspace{0.1cm}
\begin{minipage}[b]{0.26\linewidth}
\centering
\begin{tabular}{l | l c l}
$\boldsymbol\vee$ &  $\top$ & $p_0$ & $\bot$ \\
\hline
$\top$ &  $\top$ & $p_0$ & $\bot$ \\
$p_1$ &  $p_1$ & $p_0 \!\!\cross\!\! p_1$ & $\bot$ \\
$\bot$ &  $\bot$ & $\bot$ & $\bot$
\end{tabular}
\end{minipage}
\hspace{0.1cm}
\begin{minipage}[b]{0.26\linewidth}
\centering
\begin{tabular}{l | l c l}
$\boldsymbol\%$ &  $\top$ & $p_0$ & $\bot$ \\
\hline
$\top$ &  $\bot$ & $\bot$ & $\bot$ \\
$p_1$ &  $\top$ & $p_0 \!\!\cross\!\! p_1$ & $\bot$ \\
$\bot$ &  $\top$ & $\top$ & $\bot$
\end{tabular}
\end{minipage}
\hspace{0.1cm}
\begin{minipage}[b]{0.12\linewidth}
\centering
\begin{tabular}{c | c}
$\boldsymbol\neg$ & \\
\hline
$\top$ & $\bot$ \\
$p$ & $p$ \\
$\bot$ & $\top$
\end{tabular}
\end{minipage}
\label{table:composition}
\end{table}

\normalsize

We say $(I_0,O_0)$ and $(I_1,O_1)$ are \emph{$\wedge$- and $\vee$-composable} if $(I_0,O_0) = (I_1,O_1)$, and are \emph{$\%$-composable} if $(I_0,O_0)$ \emph{dominate} $(I_1,O_1)$, i.e. $A_1 \subseteq A_0$ and $O_1 \subseteq O_0$. Then, we can define the alphabet composition operations under the respective composability restriction: $(I_0,O_0) = (I_0,O_0) \wedge (I_1,O_1)$, $(I_0,O_0) = (I_0,O_0) \vee (I_1,O_1)$ and $(I_0 \cup O_1, O_0\setminus O_1) = (I_0,O_0) \% (I_1,O_1)$. 


\paragraph{Remark} Note the subtlety in the transition rules of $\P_0 \prod_{\wedge} \P_1$ and $\P_0 \prod_{\vee} \P_1$. If we have $p_0 \ar{\alpha} p'_0$ in $\P_0$ and $p_1 \ar{\alpha} \top$ in $\P_1$, then we have $p_0 \cross p_1 \ar{\alpha} p'_0$ in $\P_0 \prod_{\wedge} \P_1$. That is, process $\P_1$ is discarded after the transition and the rest of the execution is the solo run of $\P_0$.

\

Like $\prod_{\parallel}$, the definition of $\prod_{\wedge}$ can be extended without modification to work on $\bot$-complete TIOTSs (cf Footnote~\ref{fn:vee}). On specifications, $\prod_{\wedge}$ preserves the $\top$-freedom but not semi-$\top$ freedom. Thus $\P \prod_{\wedge} \Q$ may contain semi-$\top$ and has to be converted to a specification.  


In contrast, the definitions of $\prod_{\vee}$ and $\prod_{\%}$ do not extend to $\bot$-complete TIOTSs. We have to perform $\top$-completion on the operands. Then $\P^{\top} \prod_{\vee} \Q^{\top}$ and $\P^{\top} \prod_{\%} \Q^{\top}$ produce a general TIOTS, which needs to be converted back to a realisable one.

The rationale here is that the $\prod_{\vee}$, $\prod_{\wedge}$ and $\prod_{\%}$ operators implement the desiderata using the $\semb{\cdot}$ semantics rather than the $\semb{\cdot}_r$ one. Thus, $\P \prod_{\wedge} \Q$ implements $\semb{\P^{\top}} \cap \semb{\Q^{\top}}$ rather than $\semb{\P}_r \cap \semb{\Q}_r$.

\paragraph{Example} Let $\P$ be a specification that waits exactly 3 time units before firing output $a$, while $\Q$ is a specification that waits silently forever. Both are characterised by their sets of realisable strategies. However, if $\P$ and $\Q$ are put into conjunction using $\prod_{\wedge}$, then there is no realisable strategy in the intersection $\semb{\P^{\top}} \cap \semb{\Q^{\top}}$ even though the intersection is non-empty. 

\

However, it is interesting to observe that $\semb{\P}_r \cap \semb{\Q}_r = RG(\semb{\P^{\top}} \cap \semb{\Q^{\top}})$ holds for normalised specifications $\P$ and $\Q$, where the \emph{realisability filtering} function $RG(\Gamma)$ extracts the subset of realisable strategies from $\Gamma$. Thus, our conversion aims to implement the realisability filtering on top of TIOTSs.



There are two cases for such a conversion. In the first case when the resultant TIOTS is free of auto-$\top$ and semi-$\top$, $\bot$-removal suffices to remove unrealisability. This is the case for $\P^{\top} \prod_{\vee} \Q^{\top}$ since $\prod_{\vee}$ preserves the auto-$\top$ and semi-$\top$ freedom on $\top$/$\bot$-complete TIOTSs.

In the second case when the resultant TIOTS contains auto-$\top$ and semi-$\top$ (the case for $\P \prod_{\wedge} \Q$ and $\P^{\top} \prod_{\%} \Q^{\top}$), we need a more sophisticated procedure for unrealisability removal. Let us start with a deeper analysis of auto-$\top$ and semi-$\top$.

\paragraph{Auto-$\top$ and semi-$\top$ as $\top$-winning states}
Like auto-$\bot$ and semi-$\bot$, it is best to understand auto-$\top$ and semi-$\top$ in terms of perfect-information games (as determinisation does not preserve auto-$\top$ and semi-$\top$).



In a perfect-information game $\P^D$, a key observation is that \emph{a plain state $p$ is an auto-$\top$ or semi-$\top$ implies no strategy starting from $p$ is realisable}. 

For instance, if $p$ is an auto-$\top$, $p$ has an input transition going to $\top$. Then all strategies starting from $p$ have to unfold that input transition (due to determinism) and thus are unrealisable. 

If $p$ is, on the other hand, a semi-$\top$, any strategy starting from $p$, if realisable, has to make a delay move at $p$ (since all output moves lead to $\top$ due to the semi-$\top$). However, according to our strategy definition, after the delay move, which has to be finite, the strategy will have to make an output move, which unavoidably leads to $\top$.



Auto-$\top$ and semi-$\top$ characterise only a subclass of those plain states from which there is no realisable strategy. The characterisation of the full class requires, surprisingly, a dual game of the $\bot$-reachability game.

Given a perfect-information game $\P^D$ in which the collaboration of the \emph{environment and coin} play against the \emph{component} for the objective of $\top$-reachability, we say a (realisable) environment strategy $\R_E$ and a coin strategy $h \in H$ is \emph{winning} in game $\P$ (or winning against strategy set $[\P]$) iff $\R_E \cross_h \G$ can reach $\top$ for all $\G \in [\P]$. Then we say a plain state $p$ in $\P^D$ is \emph{$\top$-winning} iff there is a pair of (realisable) environment and coin strategies winning in game $\P^D(p)$. 


\paragraph{Remark} Note that $\top$- and $\bot$- winning states are dual to each other, and it is possible that a state in a TIOTS is $\bot$-winning and $\top$-winning simultaneously. However, the theory in this paper uses only a restricted class of TIOTSs, in which it is impossible to be simultaneously $\bot$-winning and $\top$-winning.

\

It is easy to verify that semi-$\top$ and auto-$\top$ are both $\top$-winning states and that the absence of semi-$\top$ and auto-$\top$ implies the absence of $\top$-winning states. 

\begin{lemma}
\label{lem:w2astop}
A TIOTS is free of $\top$-winning states iff it is free of semi-$\top$ and auto-$\top$.
\end{lemma}

Based on $\top$-winning states, we can derive a procedure (dual to normalisation) to filter out unrealisable strategies for any TIOTS.


\paragraph{Extracting realisable strategies (realisation)} Given a $\top/\bot$-complete TIOTS $\P$, using a three-step procedure we can extract the realisable subsystem $\P^R$ of $\P$ (called the \emph{realisation} of $\P$). $\P^R$ contains precisely the realisable strategies in $\semb{\P}$, i.e. $\semb{\P^R}_r=RG(\semb{\P})$.


The \emph{first step} determinises $\P$ and makes all strategies explicit. Then the \emph{second step} find and \emph{replace with $\top$} all the $\top$-winning states in $\P^D$. Finally the \emph{last step} performs a $\top$-removal on the resultant TIOTS (if it is not already the $\top$-TIOTS).








\paragraph{$\top$-backpropagation} The alternative localised approach to generating $\P^R$, called \emph{$\top$-backpropagation}, repeatedly collapses semi-$\top$ and auto-$\top$ states in $\P^D$ to $\top$ until semi-$\top$ and auto-$\top$ freedom is obtained. 

\




Hence, $\P^R$ produces either the \emph{unrealisable specification} (i.e. the $\top$-TIOTS) or a (deterministic) specification. If we define $\semb{\P^R}_r = \{\}$ for the unrealisable specification, then we have the lemma below.


\begin{lemma}
For a $\top/\bot$-complete TIOTS $\P$, $RG(\semb{\P}) = \semb{\P^R}_r$.
\end{lemma}

\paragraph{Operator definitions}
Given normalised specifications $\P$ and $\Q$, we define $\P \vee \Q$ to be the $\top$-removal of $\P^{\top} \prod_{\vee} \Q^{\top}$ and define $\P \wedge \Q = (\P \prod_{\wedge} \Q)^R$ and $\P \% \Q = (\P^{\top} \prod_{\%} \Q^{\top})^R$. The mirror operation, $\P^{\neg}$, can be defined as performing an \emph{I/O switch operation} on $\P^{\top}$, i.e. $\P^{\neg}$ is the $\top$-removal of $(\P^{\top})^X$. The I/O switch operation $\Q^X$ interchanges the input and output sets, as well as the $\top$ and $\bot$ states on $\top/\bot$-completed $\Q$.

\

Based on the mirror operator, we can give an alternative definition of quotient as the derived operator $(\P_0^{\neg}  \parallel \P_1)^{\neg}$. This is a lifting of the derivation of quotient from mirror and parallel on the state level.

Finally, we can verify that the above operator definitions implement the desiderata.

\begin{theorem}
Given a pair of $\otimes$-composable normalised specification $\P$ and $\Q$ with $\otimes \in \{\wedge, \vee, \%\}$, we have $\semb{\P \otimes \Q}_n =  \semb{\P}_n \otimes \semb{\Q}_n$ and $\semb{\P^{\neg}}_n= \semb{\P}_n^{\neg}$. 
\end{theorem}

\subsection{General case}
\label{sec:op}

For the general case when the specifications are not normalised, there is a naively correct definitions by the application of a three-step recipe. We start with normalisation, go on with applying the corresponding $\prod_{\otimes}$ operators, and finish with realisation.

However, this approach sheds little light on understanding the composability of synthesis games under the set of operators and may potentially introduce unnecessary cumbersome steps in the operator definitions. For instance, $\parallel$ is defined above without any need of normalisation or realisation. We can verify the natural definition is equivalent to the three-step recipe definition.

\begin{lemma}
Given specifications $\P$ and $\Q$, $\P \parallel \Q$ gives rise to a specification realisably equivalent to $((\P^N)^{\top} \prod_{\parallel} (\Q^N)^{\top})^R$.
\end{lemma}

The proof of the above lemma is based on the composability of normalisation games under the parallel operator, i.e. the distributivity of normalisation operation over parallel composition.

\begin{lemma}
\label{lem:18}
$(\P \parallel \Q)^D = \P^D \parallel \Q^D$ and $(\P \parallel \Q)^N = (\P^N \parallel \Q^N)^N$.
\end{lemma}

\

\begin{lemma}
\label{lem:19}
Given specifications $\P$ and $\Q$, for any product state $p \cross q$ in $\P^D \parallel \Q^D$, $p$ (or $q$) is a $\bot$-winning state in $\P^D$ (or $\Q^D$) implies $p \cross q$ is a $\bot$-winning state in $\P^D \parallel \Q^D$.
\end{lemma}

Then we can formally show that $\parallel$-composition implements strategy composition.

\begin{proposition}
For any pair of $\parallel$-composable specification $\P$ and $\Q$, we have $\semb{\P \parallel \Q}_n =  (\semb{\P}_n \cross \semb{\Q}_n)^N$.
\end{proposition}

\paragraph{Disjunction} Like the parallel operator $\parallel$, disjunction $\vee$ is also (nearly) a natural operator to define. 


\begin{lemma}
Given specifications $\P$ and $\Q$, $\P \vee \Q$ gives rise to a specification realisably equivalent to $((\P^N)^{\top} \prod_{\vee} (\Q^N)^{\top})^R$.
\end{lemma}


The proof of the above lemma is based on the composability of normalisation games under disjunction.

\begin{lemma}
\label{lem:22}
$(\P \vee \Q)^D = \P^D \vee \Q^D$ and $(\P \vee \Q)^N = (\P^N \vee \Q^N)^N$. 
\end{lemma}



\begin{lemma}
\label{lem:23}
Given specifications $\P$ and $\Q$, for any product state $p \cross q$ in $\P^D \vee \Q^D$, $p$ (or $q$) is a $\bot$-winning state in $\P^D$ (or $\Q^D$) implies $p \cross q$ is a $\bot$-winning state in $\P^D \vee \Q^D$.
\end{lemma}

The natural definitions will also work for hiding and renaming since like $\prod_{\parallel}$ and $\prod_{\vee}$ they do not generate new $\top$-winning states, although they do generate new $\bot$-winning states.

\

However, for conjunction $\wedge$ and quotient $\%$, natural definitions do not work. This is due to the subtle interferences the composition imposed on the $\top$- and $\bot$- winning states in their operands.



\begin{figure}[t]
\begin{center}
\includegraphics[width=0.9\textwidth]{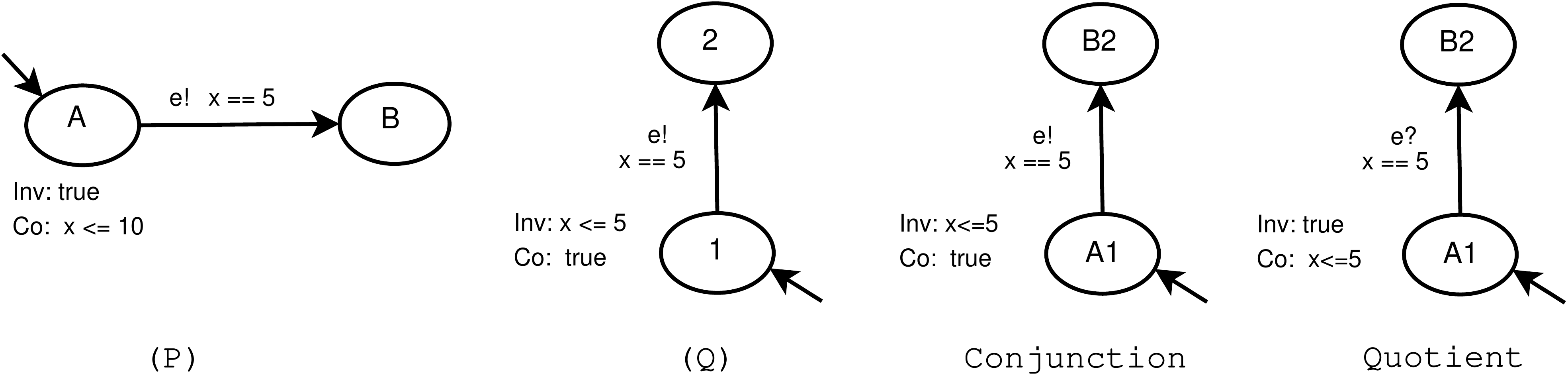}
\end{center}
\caption{Inter-component interference on winning states.}
\label{fig:CEconj}
\end{figure}

\paragraph{Example} In Figure~\ref{fig:CEconj}, we have two specifications $\P$ and $\Q$. $\Q$ is normalised while $\P$ is not. Normalisation will reduce $\P$ to the $\bot$-TIOTS (simply denoted $\bot$). It is easy to see that $\P \prod_{\wedge} \Q$ (cf~\ref{app:a}) produces the third specification, which is a normalised specification, rather than the $\bot$-TIOTS (according to $\bot \prod_{\wedge} \Q = \bot$).  This is due to the fact that with conjunction composition the $\bot$-winning states at location $A$ of $\P$ are interfered and annulled by the urgency requirement on output $e$ at location $1$ of $\Q$. Similarly, $\P \prod_{\%} \Q$ (cf~\ref{app:a}) produces the fourth specification, which is a normalised specification, rather than the $\bot$-TIOTS. 

\paragraph{Conjunction} Technically speaking, conjunction will cause interferences on the $\bot$-winning states of its operands, which leads to the non-distributivity of normalisation over $\prod_{\wedge}$, i.e. $(\P \prod_{\wedge} \Q)^N = (\P^N \prod_{\wedge} \Q^N)^N$ does not necessarily hold. 
Conjunction will not cause interferences on the $\top$-winning states of its operands though. This, combined with the distributivity of determinisation over $\prod_{\wedge}$, gives rise to distributivity of realisation over $\prod_{\wedge}$.

\begin{lemma}
Given two $\top/\bot$-complete TIOTSs $\P$ and $\Q$, we have $(\P \prod_{\wedge}$ $ \Q)^D = \P^D \prod_{\wedge} \Q^D$ and $((\P^R)^{\top} \prod_{\wedge} (\Q^R)^{\top})^R = (\P \prod_{\wedge} \Q)^R$.
\end{lemma}

Furthermore $\prod_{\wedge}$ preserves the freedom of $\bot$-winning states but not the freedom of $\top$-winning states.

\begin{lemma}
\label{lem:25}
Given two $\top/\bot$-complete TIOTSs $\P$ and $\Q$, $\P^D$ and $\Q^D$ are free of $\bot$-winning states implies $\P^D \prod_{\wedge} \Q^D$ is free of $\bot$-winning states. For any product state $p \cross q$ in $\P^D \prod_{\wedge} \Q^D$, $p$ (or $q$) is a $\top$-winning state in $\P^D$ (or $\Q^D$) implies $p \cross q$ is a $\top$-winning state in $\P^D \prod_{\wedge} \Q^D$. 
\end{lemma}

Hence, we use the three-step recipe to define conjunction. Given specifications $\P$ and $\Q$, we define $\P \wedge \Q = ((\P^N)^{\top}  \prod_{\wedge} (\Q^N)^{\top})^R$. Lemma~\ref{lem:25} implies that $\P \wedge \Q$ is a normalised specification.

For mirror and quotient, we use only part of the three-step recipe, since some transformations in the recipe are not essential for interference cancellation.

\paragraph{Mirror} The mirror of a specification $\P$, denoted $\P^{\neg}$, is defined by equation $\P^{\neg} = (((\P^{\top})^D)^X)^R$. That is, no normalisation is needed on the operand. This is because the I/O switch operation $\N^X$  (as defined in Section~\ref{sec:opsem}), rather than causing interferences on $\top$- and $\bot$- winning states in $\N$, only causes a switch between the two types of winning states. Thus, $\P^{\neg}$ is equivalent to the three-step recipe definition, i.e. the $\top$-removal of $((\P^N)^{\top})^X$. Since $\P$ as a specification is free of auto-$\top$ and semi-$\top$, $\P^{\neg}$ gives rise to a specification that is free of auto-$\bot$ and semi-$\bot$, i.e. a normalised specification.


\begin{lemma}
Given any specification $\P$, $\P^{\neg}$ is a normalised specification realisably equivalent to the $\top$-removal of $((\P^N)^{\top})^X)$.
\end{lemma}




The lemma below is very useful, since it shows how mirror can reduce the problem of refinement checking between two open systems to a non-reachability problem on a closed system.

\begin{proposition}
For any specification $\P$ and $\Q$, $\P \sqsubseteq_r \Q$ iff $\P^{\neg} \parallel Q$ is $\bot$-free.
\end{proposition}

\paragraph{Quotient} Given specifications $\P$ and $\Q$, we define $\P \% \Q = ((\P^N)^{\top} \prod_{\%} (\Q^D)^{\top})^R$. The crucial point here is that we do not need to normalise $\Q$ (i.e. the plant in the controller synthesis framework). The definition can be shown to be consistent with the one using the three-step recipe.

\begin{lemma}
Given any specification $\P$ and $\Q$, $\P \% \Q$ is a normalised specification realisably equivalent to $((\P^N)^{\top} \prod_{\%} (\Q^N)^{\top})^R$.
\end{lemma}

The proof of the above lemma is based on the composability of an order pair of normalisation and realisation games under quotient. 


\begin{lemma}
\label{lem:29}
Given two deterministic $\top/\bot$-complete TIOTSs $\P$ and $\Q$, $\P$ is free of $\bot$-winning states and $\Q$ free of $\top$-winning states implies 1) $\P \prod_{\%} \Q$ is free of $\bot$-winning states, 2) $(\P^R \prod_{\%} \Q^N)^R = (\P \prod_{\%} \Q)^R$ and 3) for any product state $p \cross q$ in $\P \prod_{\%} \Q$, $p$ is a $\top$-winning state in $\P$ or $q$ is a $\bot$-winning state in $\Q$ implies $p \cross q$ is a $\top$-winning state in $\P \prod_{\%} \Q$.
\end{lemma}


We can verify that $\P_0 \% \P_1$ gives rise to a normalised specification realisably equivalent to $(\P_0^{\neg} \parallel \P_1)^{\neg}$.

\begin{figure}[t]
\begin{center}
\includegraphics[width=\textwidth]{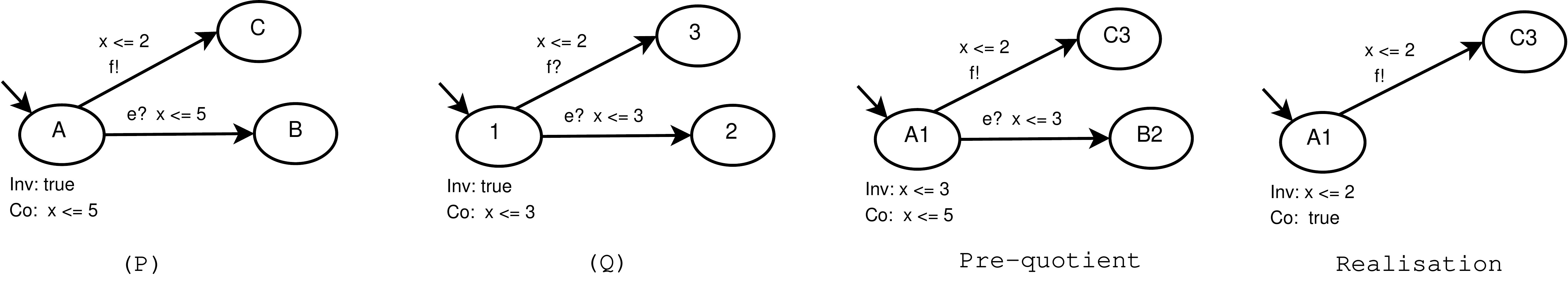}
\end{center}
\caption{Generation and removal of $\top$-winning states.}
\label{fig:topBP}
\end{figure}

\paragraph{Example} We give an example to show how $\prod_{\%}$ can generate new $\top$-winning states and how realisation can remove them. In Figure~\ref{fig:topBP}, $\P$ and $\Q$ are both normalised specifications. At location $A$, $\P$ can choose either (behaviour A) to output $f$ during the time window 0 to 2 or (behaviour B) to wait for input $e$ until time 5, at which point, if the environment fails to supply $e$, timeout will occur. On the other hand, at location $1$, $\Q$ can choose (behaviour C) either to wait for input $f$ during time window 0 to 2 or (behaviour D) to wait for input $e$ until time 3, at which point, if the environment fails to supply $e$, timeout will occur. Obviously behaviour A should be matched to behaviour C and behaviour B to D. However, the timeout bound of behaviour D is stronger than that of B. Since it is impossible to weaken one component's input assumption by composing it with another component which has to treat the action either as input or as outside the alphabet, matching D to B generate an unrealisable behaviour in the pre-quotient $\P \prod_{\%} \Q$, which can be removed by the realisation.



\

Finally, we can formally show that the operator definitions implement the desiderata.

\begin{theorem}
Given a pair of $\otimes$-composable specification $\P$ and $\Q$ with $\otimes \in \{\wedge, \vee, \%\}$, we have $\semb{\P \otimes \Q}_n =  \semb{\P}_n \otimes \semb{\Q}_n$ and $\semb{\P^{\neg}}_n= \semb{\P}_n^{\neg}$. 
\end{theorem}

Based on the above theorem we can prove the congruence result.

\begin{theorem}
$\simeq_r$ is a congruence w.r.t. $\parallel$, $\vee$, $\wedge$ and $\%$, subject to composability.
\end{theorem}

\paragraph{Double trace semantics} In addition to the timed strategy semantics, \ref{sec:trace} also gives a double trace semantics like that in our earlier work~\cite{CKW12}.


\paragraph{Timed synthesis}

Our formulation of timed synthesis games (realisation or normalisation) recognises three players in the game, i.e. coin, component and environment. On an abstract level, the two games actually belong to the same class, in which two players with reachability objective collaborate and play against the third with safety objective. Such a game has the nice properties that it is determined and winning strategies are memoryless. (For this paper we only consider the winning states for the two-player side.) 

Our $\top$- and $\bot$- backpropagations share similarities with the classical algorithms of timed synthesis games~\cite{Asarin98,Cassez05}. Both implement some form of backward fix-point computations of winning states; both can be adapted into efficient on-the-fly algorithms~\cite{Cassez05}.

However, there are some important differences. Our auto-$\bot$ and semi-$\bot$ states are related to but not equivalent to the controllable predecessors of $\bot$ in~\cite{Cassez05}. For example, an auto-$\bot$ state will not be a controllable predecessor of $\bot$ if it has an input outgoing transition leading to a plain state. Thus, our $\top$- and $\bot$- backpropagations are strictly more aggressive than the classic algorithms in classifying winning states, since the latter cannot back-propagate through auto-$\bot$. This is crucial for our weakest congruence results.

Another advantage of the three-player formulation is that the composition of the three strategies generates a run for closed systems or a strategy for open systems, thus giving rise naturally to the strategy semantics. In contrast, the composition of the two strategies in~\cite{Cassez05} does not generate a run or strategy for the composed system.

Finally,  with three-player formulation,  we can clarify the reducibility of a timed non-reachability (i.e. safety) game to a timed reachability game. For the two-player formulation it seems such reduction is possible by exchanging the role of the system and environment and complementing the target state set~\cite{Cassez05}. However, this is not true according to the three-player formulation since a game of two players with reachability objective and one player with safety objective cannot be reduced to a game of two players with safety objective and one player with reachability objective.

\paragraph{Compositional timed synthesis}


Since a specification may involve both realisation and normalisation, The composition of specifications involves the composition of synthesis games. We now understand that 1) normalisation games are composable under parallel and disjunction, 2) realisation games are composable under conjunction and 3) an ordered pair of realisation and normalisation games are composable under quotient. 

For instance, our Lemma~\ref{lem:25} implies $((\P^R)^{\top} \prod_{\wedge} (\Q^R)^{\top})^R = (\P \prod_{\wedge} \Q)^R$,  which essentially gives us a compositional method to synthesise timed processes (cf~\cite{Filiot10} for the compositional process synthesis of the untimed case).

Based on such knowledge, when composing specifications by operator $\otimes$, we now understand that only the synthesis games composable under $\otimes$ in the specifications should be composed. The incomposable ones should be removed by performing realisation or normalisation in advance. 

\section{A Printing Example}
\label{sec:exam}

\begin{figure}
\begin{center}
\includegraphics[width=\textwidth]{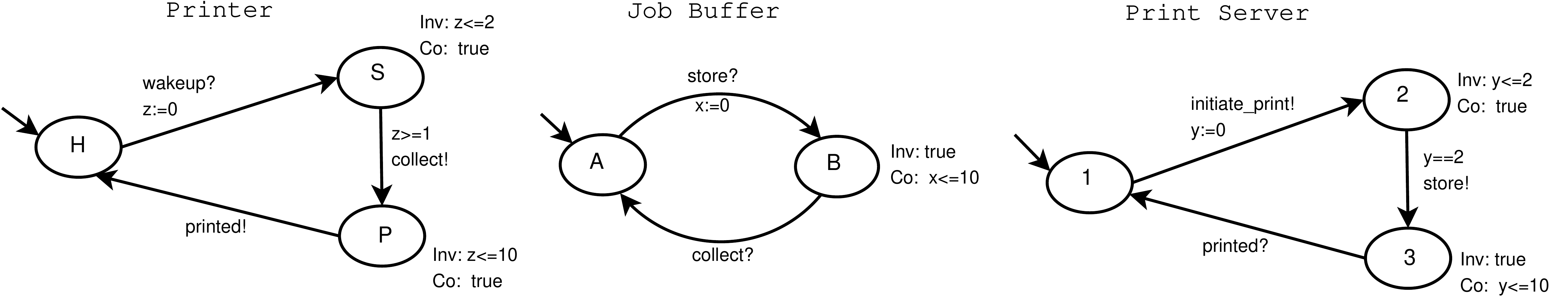}
\end{center}
\caption{Specifications for a print server, job buffer and printer.}
\label{fig:quotient}
\end{figure}

To illustrate our theory, we consider a simple printing system. Figure~\ref{fig:quotient} shows specifications of three components in the system: a print server, job buffer and printer. Intuitively, the print server decides when to $initiate\_print$ a document, after which it $store$s the job on the buffer. When the printer is told to $wakeup$, it will $collect$ the job from the buffer, and, after printing it, confirm to the print server that the job has been $printed$. The invariants, co-invariants and guards place constraints on when actions may and must occur. For example, once the printer has been told to $wakeup$, it must $collect$ a job at least 1s, although no more than 2s, later and the document must have been $printed$ within 10s, in order to satisfy the invariants. After the job buffer has been told to $store$ a job, the co-invariant requires that the job is $collect$ed within 10s. For the print server, after deciding to $initiate\_print$, the job must be $store$d exactly 2s later (imposed by the invariant and guard on state 2), and requires that the job must have $printed$ within 10s (imposed by the co-invariant on state 3).

\begin{figure}
\begin{center}
\includegraphics[width=\textwidth]{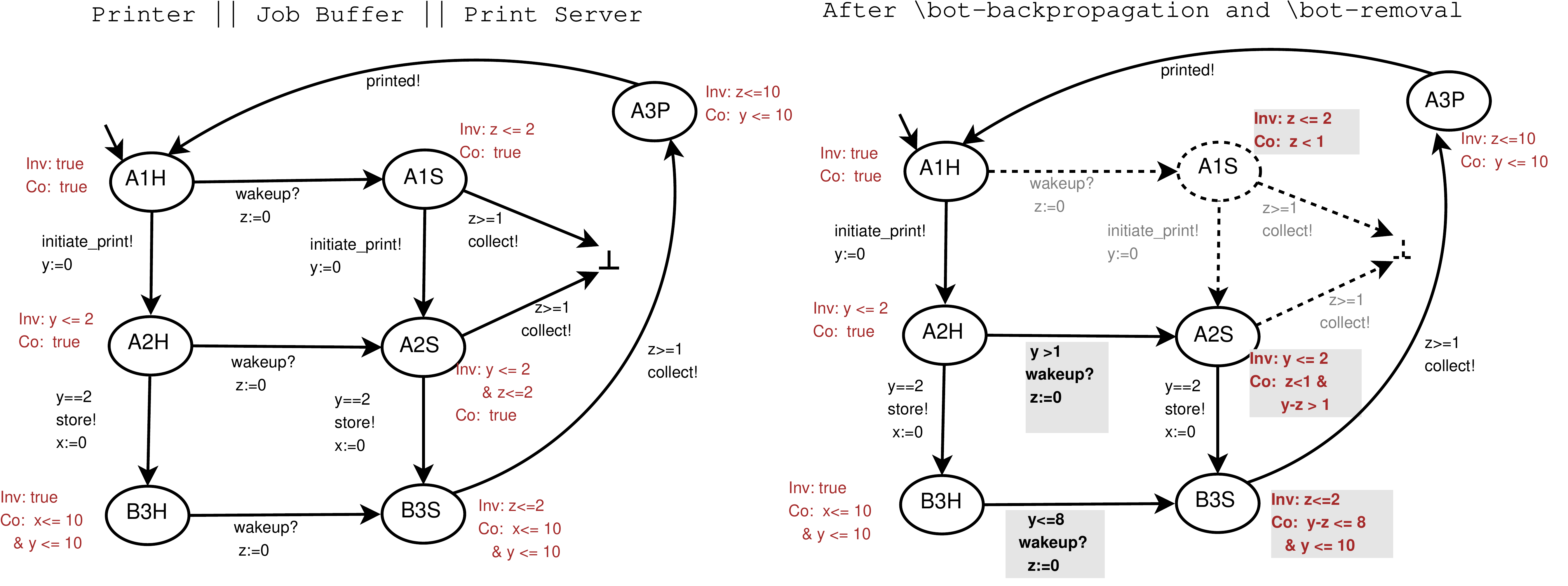}
\end{center}
\caption{Parallel composition of the print server, job buffer and printer, and $\bot$-backpropagation.}
\label{fig:synthesis}
\end{figure}

The three components can be composed under parallel. However, they will not work together without external coordination. For example, the $wakeup$ input to the printer is not supplied by any of the other two components. Thus, we need a \emph{scheduler} which can connect the three components together and produce the $wakeup$ at the right time. The clever bit here lies in the synthesis of the scheduler strategies such that the printer is not told to $wakeup$ too early or too late.


Basically, we synthesise the scheduler by calculating the least refined environment such that the three can work together without violating any of their timing constarints: $(Printer \parallel Job\_Buffer \parallel Print\_Server)^{\neg}$.

The left-hand side of Figure~\ref{fig:synthesis} shows the parallel composition of the three components in Figure~\ref{fig:quotient}, i.e. $System = Printer \parallel Job\_Buffer \parallel Print\_Server$, which is essentially the synchronised product of the specifications by taking the conjunction of invariants, co-invariants and guards. The $\bot$-state is reachable due to non-input enabledness of the $collect$ transition in the job buffer (the printer collects the job too early or too late).

To perform mirroring on $System$, it must first be normalised. We implement the normalisation by a $\bot$-backpropagation followed by $\bot$-removal on $System$.\footnote{$\bot$-removal is not strictly necessary for mirroring, but it simplifies the result for better readability.} On the right-hand side of Figure~\ref{fig:synthesis}, we show the resultant TIOA after the two transformations.

Since the output transition $collect$ at location $A1S$ leads to $\bot$, those states associated with location $A1S$ on which $collect$ is enabled will be auto-$\bot$ states. Collapsing them to $\bot$ is equivalent to strengthening the co-invariant on $A1S$ to keep only those states on which $collect$ is not enabled. Thus the co-invariant is changed to $z<1$.\footnote{Note that we use shaded areas in the right-hand side of Figure~\ref{fig:synthesis} to mark the guards and invariants/co-invariants changed by the transformations.}

After the change, however, the invariant at $A1S$ becomes redundant. Thus all the remaining states associated with $A1S$ become semi-$\bot$ states since there is no outgoing input transition at $A1S$. Thus, location $A1S$ can completely collapse to $\bot$, culminating in the removal of its associated transitions (indicated by dotted lines).

For location $A2S$, similarly its co-invariant can be changed to $z<1$ due to the auto-$\bot$ caused by its $collect$ transition. But the new co-invariant will not make its invariant completely redundant. Instead, it is only when the co-invariant can reach its upper bound before the invariant reaches its (i.e. when $y-z <= 2-1$) that the states at location $A2S$ becomes semi-$\bot$. Thus, the co-invariant needs to be changed to $y-z > 1 \& z <1$. Then we can perform $\bot$-removal on the incoming $wakeup$ transition by removing the $wakeup$ transition whose firing will make $y-z <= 1$ true. Thus, the guard $y>1$ is added to the $wakeup$ transition.

Similarly, location $B3S$ has semi-$\bot$ if $y-z>10-2$. Thus its co-invariant needs to be changed to $y-z <= 8 \& y <=10$ and its incoming $wakeup$ transition needs to be strengthened with the guard $y<=8$.

After the two transformations, we need to perform the mirror operation on the resultant TIOA by exchanging input with output and invariant with co-invariant. Then the final TIOA will be our synthesised scheduler. Due to the synthesis procedure, infeasible strategies, such as issuing $wakeup$ before receiving $initial\_print$ or issuing $wakeup$ after receiving $initial\_print$ but before clock $y$ reaching 1s, are automatically eliminated.


\section{Comparison with Related Work}
\label{sec:comp}


Our framework can be seen as a linear-time alternative to the timed specification theories of~\cite{henzinger-timedia} and~\cite{larsen-timedio}, albeit with significant differences. The specification theory in~\cite{larsen-timedio} also introduces parallel, conjunction and quotient, but uses timed alternating simulation as refinement, which does not admit the weakest precongruence (cf $P$ and $Q$ in Figure~\ref{fig:strategy-equivalence-strong}). An advantage of~\cite{larsen-timedio} is the algorithmic efficiency of branching-time simulation checking and implementation reported in~\cite{DLLNW10}.

The work of~\cite{henzinger-timedia} on timed games shares significantly more conceptual and technical similarities with us, although they do not define refinement, conjunction and quotient.
We adopt most of the game rules in~\cite{henzinger-timedia}, except that, due to our requirement that proposed delay moves are maximal delays allowed by a strategy, a play
cannot have consecutive delay moves.

This enables us to avoid the complexity of an infinite play (i.e. infinite sequence of moves) generating a finite trace (cf Section~\ref{sec:timetrans} for the definition of finite traces). So infinite plays generate only divergent traces (cf the non-zenoness assumption). To completely eliminate time-blocking strategies, we only need to tackle the remaining case that finite plays end in timestop or timelock, which can be nicely solved using the realisation game. Thus the need for blame assignment is removed.

Secondly, we do not use timelock (i.e. semi-$\top$) to model time errors (i.e. bounded-liveness errors). Rather, we introduce the explicit inconsistent state $\bot$ to model both time and immediate (i.e. safety) errors. This enables us to avoid the complexity of having two transition relations and well-formedness of timed interfaces. 

Similar to our work, \cite{larsen-timedio} uses semi-$\top$ to model timelock (so-called \emph{immediate errors} in~\cite{larsen-timedio}). However, the pruning of timelocks is based on the synthesis game of~\cite{Cassez05}. Therefore, they cannot remove auto-$\top$ and the pruning is strictly less aggressive. 

Furthermore, incompatibility errors (so-called \emph{strictly undesirable states} in~\cite{larsen-timedio}) are not in the core of the theory for~\cite{larsen-timedio}. They are more `model-related errors' defined by the users, which are treated as plain states by the definition of operators and refinement. So it is unclear (e.g. for conjunction and qotient) what the product state will be if one component is in strictly undesirable states. 

This is in contrast to our theory, where the definition of the four operators, substitutive refinement relations, and determinisation procedure are all based on the manipulation of $\top$ and $\bot$; and the algebraic properties from state composition operators can be lifted to the process level. 

More specifically, some further technical points of comparison with~\cite{larsen-timedio,henzinger-timedia} are:

\begin{itemize}
\item \emph{Determinism:} We can handle non-deterministic timed transition systems thanks to our modified determinisation procedure while~\cite{larsen-timedio,henzinger-timedia} consider only deterministic timed transition system. That is where a linear time theory have advantages. It is not obvious how such extension can work if the refinement is timed alternating simulation. 

\item \emph{AG reasoning:} A specification in~\cite{larsen-timedio} is an input-enabled TIOA/TIOTS without $\bot$ or co-invariants. Thus a specification contains no assumptions on the environment before users mark out strictly undesirable states. It is not a fully assume-guarantee specification theory in the sense that a specification (or interface) combines and mixes assumptions and guarantees in a unified way.


\item \emph{Implementation and strategy:} A specification in~\cite{larsen-timedio} can be interpreted as a set of implementations while our timed strategy semantics interprets a specification as a set of strategies. There is some similarity. However, the major differences are:

\begin{itemize}
\item Strategies are tree-like partial unfoldings of original transition system while implementation are (potentially cyclic) transition systems alternating simulating the original system.

\item We have implicit strategies which can be neither partial unfoldings nor alternating simulation of the original systems.

\item Strategies are based on game theory and use game rules like those in~\cite{henzinger-timedia}. However, implementation is less closely related to game theory. 
\end{itemize}



\end{itemize}


In comparison with the untimed specification theories~\cite{ESOP}, our timed extension requires new techniques (e.g. those related to timestop) to handle delay transitions since time can be modelled neither as input nor as output. Timestop enables us to discover the surprisingly simple and robust notions like semi-$\top/\bot$ and $\top/\bot$-backpropagation, whose definitions indicate the canonicity of the notions. Furthermore, with the assistance of time, bounded liveness in terms of clock bounds suffices to specify and verify most liveness-related properties. Bounded liveness is especially simple and natural to use and work with in timed models since invariant/co-invariant and finite traces suffice to capture. In contrast, in the untimed world, bounded liveness is cumbersome to specify and work with; people in most cases have to resort to infinite traces to treat liveness properly.

Finally, we remark that our linear-time specification theory owes much to the pioneering work on trace theories for asynchronous circuit verification, such as Dill's trace theory~\cite{dill-trace-theory}. It is from this community that we take inspiration for the timed extension of mirror and the derivation of quotient from mirror\footnote{The mirror-based definition of quotient (for the untimed case) was first presented by Verhoeff as his Factorisation Theorem~\cite{verhoeff-thesis}.}. In some sense, this work can be regarded as a combination of this line of work with another line of work to which Dill has also made the seminal contribution, timed automata. It is highly satisfying to see the synergy between the two lines of works, as indicated by the results in this work.


We briefly mention other related works, which include timed modal transition systems~\cite{BLPR09,Cerans93}, the timed I/O model~\cite{Kaynar,BV08} and embedded systems~\cite{Thiele06,Lee07}.

\section{Conclusion and Future Work}
\label{sec:concl}

We have devised a fully compositional specification theory for realisable components with real-time constraints. The linear-time theory enjoys strong algebraic properties, supports a full set of composition operators, and admits the weakest substitutive pre-congruence preserving safety and bounded-liveness error freedom.
The framework can be seen as an alternative to, or refinement of, the timed theories of~\cite{henzinger-timedia,larsen-timedio}. Future work will consider assume-guarantee reasoning for timed systems, as well as the implementation of our theory. The latter, we believe, can benefit from the timed-game based algorithms and results from~\cite{larsen-timedio}.


\paragraph{Acknowledgments} The authors are supported by EU FP7 project CONNECT, ERC Advanced Grant VERIWARE and EPSRC project EP/F001096.

\appendix

\section{Composing TIOA}
\label{app:a}



We use $\otimes$ to range over the operator set $\{\parallel, \vee, \wedge, \% \}$, and use $l$ and $n$ to range over the set of locations (i.e. $L$). 

We say a TIOA, $\P = (C, I, O, L, n^0, AT, Inv, coInv)$, is $\top$-completed iff, for all $a \in O$ and $l \in L$, we have $\bigvee \{g_k | l \ar{g_k,a,rs_k} l'_k \in T \} =true$. Note that, unlike the definition for TIOTSs, TIOAs do not require $\top$-completion on delay transitions. We say $\P$ is $\bot$-completed iff, for all $a \in I$ and $l \in L$, we have $\bigvee \{g_k | l \ar{g_k,a,rs_k} n_k \in T \} =true$.

Given two $\otimes$-composable $\top/\bot$-completed TIOAs with disjoint clocks ($C_0 \cap C_1 = \{\}$), $\P_i = (C_i, I_i, O_i, L_i, n^0_i, AT_i, Inv_i, coInv_i)$ for $i \in \{0,1\}$, their synchronised product gives rise to another TIOA $\P = \P_0 \prod_{\otimes} \P_1$:

\begin{itemize}
\item $C = C_0 \cup C_1$, $(I,O) = (I_0, O_0) \otimes (I_1, O_1)$ and $L = L_0 \times L_1$;
\item  $n^0 = n^0_0 \cross n^0_1$;
\item $AT$ is the least relation that contains $AT_0$, $AT_1$ and $\{ l_0 \cross l_1 \ar{g_0 \land g_1, a, rs_0 \cup rs_1} n'_0 \cross  n'_1  | l_0 \ar{g_0,a,rs_0} n'_0 \in AT_0 \land l_1 \ar{g_1,a,rs_1} n'_1 \in AT_1 \} \\
\cup \{l_0 \cross l_1 \ar{g_0, a, rs_0} n'_0 \cross l_1  | l_0 \ar{g_0,a,rs_0} n'_0 \in AT_0, a \in (A_0 \setminus A_1) \} \\
\cup \{l_0 \cross l_1 \ar{g_1, a, rs_1} l_0 \cross n'_1  | l_1 \ar{g_1,a,rs_1} n'_1 \in AT_1, a \in (A_1 \setminus A_0) \}  \}$;
\item and $(Inv(l_0 \cross l_1), coInv(l_0 \cross l_1)) = (Inv_0(l_0), coInv_0(l_0)) \otimes (Inv_1(l_1), coInv_1(l_1))$.
\end{itemize}

We define the $\otimes$ invariant/co-invariant composition operation as follows:

\begin{itemize}
\item $(Inv_0, coInv_0 )  \parallel (Inv_1, coInv_1 ) = (Inv_0 \land Inv_1, coInv_0 \land coInv_1)$

\item $(Inv_0, coInv_0 )  \wedge (Inv_1, coInv_1 ) = (Inv_0 \land Inv_1, coInv_0 \lor coInv_1)$

\item $(Inv_0, coInv_0 )  \vee (Inv_1, coInv_1 ) = (Inv_0 \lor Inv_1, coInv_0 \land coInv_1)$

\item $(Inv_0, coInv_0 )  \% (Inv_1, coInv_1 ) = (Inv_0 \land coInv_1, coInv_0 \land Inv_1)$
\end{itemize}

Note that in the above definition we exploit the fact that the addition or removal of $false$-guarded transitions to $AT$ will not change the semantics of the automata.

Strongly non-zeno TAs are known to be determinisable. For instance, \cite{Bertrand} gives a symbolic procedure based on game and region construction. We can easily modify the procedure to implement the TIOTS determinisation defined in Section 2, giving rise to the new procedure $DET(\P)$ on TIOA $\P$.

On deterministic TIOAs, we can implement both $\top$- and $\bot$- backpropagation procedures by fixpoint calculation on top of constraint backpropagation, denoted as $BP(\P,\top)$ and $BP(\P,\bot)$ resp.

With such transformations on TIOAs, all the operators in theory I and II become definable on TIOAs from the $\prod_{\otimes}$ operators on TIOAs.

\section{Declarative Theory of Contracts}
\label{sec:trace}

We now present a timed-trace characterisation of our compositional specification theory. For this purpose we adopt the \emph{contract} framework promoted in~\cite{Benveniste}, which has the advantage of explicitly separating \emph{assumptions} from \emph{guarantees}.

Given any TIOTS $\P= \langle I, O, S, s^0, \rightarrow \rangle$, three sets of traces can be extracted from $((\P^{\bot})^{\top})^D$:
\begin{itemize}
\item $TP$ a set of timed traces leading to plain states
\item $TE$ a set of timed traces leading to the error state $\bot$
\item $TM$ a set of timed traces leading to the magic state $\top$.
\end{itemize}
$TE$ and $TM$ are extension-closed due to the chaotic nature of $\top$ and $\bot$, while $TP$ is prefix-closed. Since $TE \cup TP \cup TM$ is the full set of timed traces (i.e. $tA^*$), we need only two of the trace sets to characterise $\P$.

In the system-environment interaction (as explained in our timed game framework), $TE$ is the set of behaviours which the environment tries to steer the interaction away from, whereas $TM$ is the set of behaviours which the component tries to steer away from. Thus, $TE$ characterises the assumptions required on the environment while $TM$ characterising the guarantees provided by the system. 


A \emph{contract} based on $TE$ and $TM$ defines the semantics of $\P$, characterising the congruence $\simeq$~\cite{CKW12}.

\begin{definition}[Contract]
A \emph{contract} is a tuple $(I, O, AS, GR)$, where $AS$ and $GR$ are two disjoint extension-closed trace sets. The contract of $\P$ is defined as $\TT{\P} := (I, O, TE, TM)$.
\end{definition}

\noindent




When $\P$ is a specification (including the unrealisable specification\footnote{When $\P$ is the unrealisable specification, i.e. the $\top$-TIOTS, $\overline{GR}$ is empty.}), $\overline{GR}$ in $\TT{\P}$ is \emph{I-receptive}. We say a trace set $TT$ is \emph{I-receptive} iff, for each $tt \in TT$, we have 1) $tt \cat \langle e \rangle \in TT$ for all $e \in I$ and 2)  $tt \cat \langle d \rangle \notin TT$ for some $d \in \mathbb{R}^{>0}$ implies there exists $w \in tO^*$ s.t. $tt \cat w \in TT$ and $l(w) < d$. 

When $\P$ is a normalised specification (including the inconsistent specification\footnote{When $\P$ is the inconsistent specification, i.e. the $\bot$-TIOTS, $\overline{AS}$ is empty.}), we have furthermore that $\overline{AS}$ in $\TT{\P}$ is \emph{O-receptive}. We say a trace set $TT$ is \emph{O-receptive} iff, for each $tt \in TT$, we have 1) $tt \cat \langle e \rangle \in TT$ for all $e \in O$ and 2) $tt \cat \langle d \rangle \notin TT$ for some $d \in \mathbb{R}^{>0}$ implies there exists $w \in tI^*$ s.t. $tt \cat w \in TT$ and $l(w) < d$.

Given a TIOTS $\P$, the realisation of $\P$, i.e. $\P^R$, can be implemented by $\top$-backpropagation on contracts:

\begin{definition}[Realisation]
Given a contract $(I, O, AS, GR)$, we define $(I, O, AS, GR)^R = (I, O, AS \setminus GR^R, GR^R)$, where $GR^R$ is the least extension-closed superset of $GR$ s.t. no $tt \in tA^*$ is an auto-$\top$ or semi-$\top$ w.r.t. $GR^R$.
\end{definition}

We say a trace $tt \in tA^*$ is an \emph{auto-$\top$} w.r.t. $TT$ iff $tt \notin TT$ and $tt \cat \langle e \rangle \in TT$ for some $e \in I$. A trace $tt \in tA^*$ is an \emph{semi-$\top$} w.r.t. $TT$ iff $tt \notin TT$ and there exists some $d \in \mathbb{R}^{>0}$ s.t. $tt \cat \langle d \rangle \in TT$ and $tt \cat \langle d_0, e \rangle \in TT$ for all $0 \leq d_0 < d$ and $e \in O$. It is easy to verify $\overline{GR^R}$ is I-receptive and $\TT{\P}^R=\TT{\P^R}$.

Given a specification $\P$, the normalisation of $\P$, i.e. $\P^N$, can be also implemented by $\bot$-backpropagation on contracts:

\begin{definition}[Normalisation]
Given a contract $(I, O, AS, GR)$ with I-receptive $\overline{GR}$, we define $(I, O, AS, GR)^N = (I, O, AS^N, GR \setminus AS^N)$, where $AS^N$ is the least extension-closed superset of $AS$ s.t. no $tt \in tA^*$ is an auto-$\bot$ or semi-$\bot$ w.r.t. $AS^N$.
\end{definition}

A trace $tt \in tA^*$ is an \emph{auto-$\bot$} w.r.t. $TT$ iff $tt \cat \langle e \rangle \in TT$ for some $e \in O$. A trace $tt \in tA^*$ is a \emph{semi-$\bot$} iff there exists some $d \in \mathbb{R}^{>0}$ s.t. $tt \cat \langle d \rangle \in TT$ and $tt \cat \langle d_0, e \rangle \in TT$ for all $0 \leq d_0 < d$ and $e \in I$. It is easy to verify that $\overline{AS^N}$ is O-receptive and $\TT{\P}^N=\TT{\P^N}$.

A coarsening of contracts gives a characterisation of $\simeq_r$, which says $\P$ is an refinement of $\Q$ iff $\P$ has less assumption and more guarantee than $\Q$.


\begin{definition}[Realisable contract] 
A contract $(I, O, AS, GR)$ is a \emph{realisable contract} iff $\overline{AS}$ is O-receptive and $\overline{GR}$ is I-receptive. The realisable contract of a specification $\P$ is defined as $\BT{\P} := \TT{\P}^N$.

\end{definition}



\begin{theorem}
\label{trace-sem}
For specifications $\P_0$ and $\P_1$ with realisable contracts $(I, O, AS_0,$ $GR_0)$ and $(I, O, AS_1, GR_1)$ respectively, $\P_0 \sqsubseteq_r \P_1$ iff $AS_1 \subseteq AS_0$ and $GR_0 \subseteq GR_1$.
\end{theorem}

Given two specifications $\P_i$ for $i \in \{0,1\}$ and $\bar{i}=1-i$ s.t. $\BT{\P_i} = (I, O, AS_i, GR_i)$, we define the parallel, disjunction, conjunction and quotient operations on realisable contracts. The core part of the operations is based on the patterns originally discovered by~\cite{dill-trace-theory,Negulescu}. The specialisation required for the timed theory to work lies in the application of closure conditions like normalisation and realisation.

We first define the alphabet enlargement operation on realisable contracts before carrying on defining the major operators.

\paragraph{Alphabet enlargement} Given a set $\Delta$ of actions disjoint from $I \cup O$, we define $(I, O, AS, GR)^\Delta := (I \cup \Delta, O, AS^{\Delta}, GR^{\Delta})$, where $TT^{\Delta} :=  \{ tt :(tA \cup \Delta)^* | tt \upharpoonright tA \in TT \}  \cdot (tA \cup \Delta)^*$. 

\paragraph{Parallel composition and disjunction}

\begin{proposition}
If specifications $\P_0$ and $\P_1$ are $\parallel$-composable, then $\BT{\P_0 \parallel \P_1} = (I, O, $ $(AS_0^{\Delta_0} \cup AS_1^{\Delta_1}) \setminus  (GR^{\Delta_0}_0 \cup GR^{\Delta_1}_1), GR^{\Delta_0}_0 \cup GR^{\Delta_1}_1)^N$, where $I=(I_0\cup I_1) \setminus O$, $O=O_0\cup O_1$, $\Delta_0 = A_1 \setminus A_0$ and $\Delta_1 = A_0 \setminus A_1$.
\end{proposition}

Intuitively, the above says that the guarantee of the parallel composition is the combined guarantees provided by the components while the assumption of the parallel composition is the combined assumptions of the components minus those that have been fulfilled by their guarantees.


\begin{proposition}
If specifications $\P_0$ and $\P_1$ are $\vee$-composable, then $\BT{\P_0 \vee \P_1} = (I, O,$ $AS_0 \cup AS_1, GR_0 \cap GR_1)^N$, where $I = I_0 = I_1$ and $O=O_0 = O_1$.
\end{proposition}

That is, disjunction unions assumptions and intersects guarantees.


\paragraph{Conjunction and quotient} 

\begin{proposition}
If $\P_0$ and $\P_1$ are $\wedge$-composable, then $\BT{\P_0 \wedge \P_1} = (I, O, $ $AS_0 \cap AS_1, GR_0 \cup GR_1)^R$, where $I = I_0 = I_1$ and $O=O_0 = O_1$.
\end{proposition}


\begin{proposition}
If specification $\P_0$ dominates specification $\P_1$, then $\BT{\P_0 \% \P_1} = (I, O, AS_0 \cup GR^{\Delta_1}_1, (GR_0 \setminus GR^{\Delta_1}_1) \cup (AS^{\Delta_1}_1 \setminus AS_0) )^R$, where $I= I_0 \cup O_1$, $O=O_0\setminus O_1$ and $\Delta_1 = A_0 \setminus A_1$.
\end{proposition}

Intuitively the above says that the quotient assumes the $\P_0$-assumption combined with the $\P_1$-guarantee and it guarantees 1) the $\P_0$-guarantee not covered by $\P_1$-guarantee as well as 2) the $\P_1$-assumption missing from $\P_0$-assumption.  




\paragraph{Mirror} The operation is straightforward, which simply exchanges assumption and guarantee.

\begin{proposition}
$\BT{\P^{\neg}} = (O, I, GR, AS)$.
\end{proposition}

\paragraph{Contract} The terminology of contract was coined by Meyer and Back. The meta-theory of contract dates back to the trace theory of~\cite{dill-trace-theory}, esp. one of its abstract reformulation by~\cite{Negulescu}. Both work draws upon earlier ideas from asynchronous circuit verification.

\bibliographystyle{plain}
\bibliography{timed-bib}

\end{document}